\documentclass[11pt,letterpaper]{article}

\IfFileExists{latexml.sty}{\usepackage{latexml}}{\newif\iflatexml\latexmlfalse}

\usepackage[T1]{fontenc}
\usepackage[utf8]{inputenc}
\iflatexml
\newcommand{\carlito}{}\newcommand{\caladea}{}
\else
\usepackage{lmodern}                 %
\usepackage{caladea}                 %
\usepackage[lining]{carlito}         %
\usepackage[default,lining]{lato}    %
\DeclareFontFamily{T1}{LatoW}{}
\DeclareFontShape{T1}{LatoW}{m}{n}{<-> Lato-Regular-T1-TLF--base}{}
\DeclareFontShape{T1}{LatoW}{m}{it}{<-> Lato-Italic-T1-TLF--base}{}
\DeclareFontShape{T1}{LatoW}{b}{n}{<-> Lato-Bold-T1-TLF--base}{}
\DeclareFontShape{T1}{LatoW}{b}{it}{<-> Lato-BoldItalic-T1-TLF--base}{}
\DeclareFontShape{T1}{LatoW}{m}{sl}{<-> ssub * LatoW/m/it}{}
\DeclareFontShape{T1}{LatoW}{b}{sl}{<-> ssub * LatoW/b/it}{}
\DeclareFontShape{T1}{LatoW}{bx}{n}{<-> ssub * LatoW/b/n}{}
\DeclareFontShape{T1}{LatoW}{bx}{it}{<-> ssub * LatoW/b/it}{}
\DeclareFontShape{T1}{LatoW}{bx}{sl}{<-> ssub * LatoW/b/it}{}
\DeclareFontFamily{TS1}{LatoW}{}
\DeclareFontShape{TS1}{LatoW}{m}{n}{<-> ssub * lato-TLF/m/n}{}
\DeclareFontShape{TS1}{LatoW}{m}{it}{<-> ssub * lato-TLF/m/it}{}
\DeclareFontShape{TS1}{LatoW}{b}{n}{<-> ssub * lato-TLF/b/n}{}
\DeclareFontShape{TS1}{LatoW}{b}{it}{<-> ssub * lato-TLF/b/it}{}
\DeclareFontShape{TS1}{LatoW}{bx}{n}{<-> ssub * lato-TLF/b/n}{}
\DeclareFontShape{TS1}{LatoW}{bx}{it}{<-> ssub * lato-TLF/b/it}{}

\DeclareFontFamily{T1}{CrltW}{}
\DeclareFontShape{T1}{CrltW}{m}{n}{<-> Crlt-tlf-t1--base}{}
\DeclareFontShape{T1}{CrltW}{m}{it}{<-> Crlt-Italic-tlf-t1--base}{}
\DeclareFontShape{T1}{CrltW}{b}{n}{<-> Crlt-Bold-tlf-t1--base}{}
\DeclareFontShape{T1}{CrltW}{b}{it}{<-> Crlt-BoldItalic-tlf-t1--base}{}
\DeclareFontShape{T1}{CrltW}{bx}{n}{<-> ssub * CrltW/b/n}{}
\DeclareFontShape{T1}{CrltW}{bx}{it}{<-> ssub * CrltW/b/it}{}
\renewcommand*{\carlito}{\fontfamily{CrltW}\selectfont}
\usepackage{textcomp}
\usepackage[protrusion=false,expansion=false]{microtype} %
\DisableLigatures[f]{encoding = *, family = *}  %
\input{glyphtounicode}\pdfgentounicode=1
\pdfglyphtounicode{fi}{0066 0069}\pdfglyphtounicode{f_i}{0066 0069}\pdfglyphtounicode{fl}{0066 006C}\pdfglyphtounicode{f_l}{0066 006C}
\pdfglyphtounicode{ff}{0066 0066}\pdfglyphtounicode{f_f}{0066 0066}\pdfglyphtounicode{ffi}{0066 0066 0069}\pdfglyphtounicode{f_f_i}{0066 0066 0069}
\pdfglyphtounicode{uniFB00}{0066 0066}\pdfglyphtounicode{uniFB01}{0066 0069}\pdfglyphtounicode{uniFB02}{0066 006C}\pdfglyphtounicode{uniFB03}{0066 0066 0069}\pdfglyphtounicode{uniFB04}{0066 0066 006C}
\pdfglyphtounicode{f_t.liga}{0066 0074}\pdfglyphtounicode{f_f_t.liga}{0066 0066 0074}\pdfglyphtounicode{f_dotlessi.liga}{0066 0131}
\fi
\DeclareUnicodeCharacter{2192}{\ensuremath{\rightarrow}}
\DeclareUnicodeCharacter{2014}{\textemdash\penalty\exhyphenpenalty}%
\DeclareUnicodeCharacter{2264}{\ensuremath{\leq}}
\DeclareUnicodeCharacter{2212}{\ensuremath{-}}

\usepackage[letterpaper,left=56.3bp,right=56.3bp,top=51.6bp,bottom=54.5bp,
            headheight=14bp,headsep=19.3bp,footskip=26.5bp]{geometry}
\usepackage[table]{xcolor}
\usepackage{graphicx}
\usepackage{array}
\usepackage{longtable}
\usepackage{enumitem}
\usepackage{titlesec}
\iflatexml\else\usepackage[titles]{tocloft}\fi
\usepackage{fancyhdr}
\usepackage{lastpage}
\usepackage{caption}
\usepackage{float}
\iflatexml\else\usepackage[absolute]{textpos}\fi
\usepackage[document]{ragged2e}
\iflatexml\else\usepackage[most]{tcolorbox}\fi
\usepackage[nocompress,nosort]{cite}  %
\usepackage{url}
\usepackage{etoolbox}
\AtBeginEnvironment{itemize}{\listsize}
\AtBeginEnvironment{enumerate}{\listsize}
\usepackage[hidelinks]{hyperref}
\hypersetup{
  pdftitle={Provable Compliance: A Protocol for Runtime Compliance Evidence and Transaction-Specific Decisioning},
  pdfauthor={Jeremy Blackburn, Tim Hines, Matthew Koepp, Russell Lessard, Jason Patrick},
  pdfsubject={ChainIT Compliance Protocol architecture and source basis (CIT-PC-WP-4.0)},
  pdfkeywords={compliance evidence, runtime verification, AML/CFT, sanctions screening, verifiable credentials, policy-based decisioning, agentic commerce, stablecoins, ChainIT}
}

\definecolor{body}{HTML}{1F2937}
\definecolor{cnavy}{HTML}{1B2A4A}
\definecolor{cblue}{HTML}{2B6CB0}
\definecolor{cgray}{HTML}{4A5568}
\definecolor{hgray}{HTML}{6B7280}
\definecolor{tgray}{HTML}{5E6C76}
\definecolor{crule}{HTML}{CBD5E0}
\definecolor{thnavy}{HTML}{1B2A4A}
\definecolor{rowalt}{HTML}{F5F7F9}
\definecolor{pblue}{HTML}{2B6CB0}\definecolor{tblue}{HTML}{EAF2F7}
\definecolor{pgreen}{HTML}{2F7D4A}\definecolor{tgreen}{HTML}{EAF4ED}
\definecolor{pgold}{HTML}{C3922E}\definecolor{tgold}{HTML}{FBF3E1}
\definecolor{pred}{HTML}{9B2C2C}\definecolor{tred}{HTML}{FBECEC}

\renewcommand{\normalsize}{\fontsize{10.5bp}{14.24bp}\selectfont}
\newcommand{\boxsize}{\fontsize{9.5bp}{12.8bp}\selectfont}
\newcommand{\listsize}{\fontsize{10bp}{13.65bp}\selectfont}
\newcommand{\smallsize}{\fontsize{9.5bp}{12.8bp}\selectfont}
\AtBeginDocument{\normalsize\color{body}}
\iflatexml\else
\setlist[itemize]{label=\textbullet,leftmargin=20.2bp,labelindent=8.7bp,labelsep=*,itemsep=0bp,parsep=0bp,topsep=3bp,after=\vspace{-1.3bp}}
\setlist[enumerate]{label=\arabic*.,leftmargin=20.2bp,labelindent=8.7bp,labelsep=*,itemsep=0bp,parsep=0bp,topsep=3bp,after=\vspace{-1.3bp}}
\fi
\newif\ifpanellast\newif\iftablelast
\iflatexml\else\AddToHook{para/begin}{\global\panellastfalse\global\tablelastfalse}\fi

\titleformat{\section}{\raggedright\hyphenpenalty=10000\exhyphenpenalty=10000\carlito\bfseries\fontsize{16bp}{19.5bp}\selectfont\color{cblue}}{\rule{0bp}{15.4bp}\thesection.}{0.25em}{\rule{0bp}{15.4bp}}
\titleformat{\subsection}{\raggedright\hyphenpenalty=10000\carlito\bfseries\fontsize{13bp}{16bp}\selectfont\color{cblue}}{\rule{0bp}{12.7bp}\thesubsection}{0.25em}{\rule{0bp}{12.7bp}}
\titlespacing*{\section}{0bp}{\ifpanellast 24.7bp\else\iftablelast 17.6bp\else 7.3bp\fi\fi}{3.3bp}
\titlespacing*{\subsection}{0bp}{\ifpanellast 19.8bp\else\iftablelast 15.3bp\else 5bp\fi\fi}{1bp}
\newcommand{\minorhead}[1]{\par\vspace{\ifpanellast 5.3bp\else\iftablelast 5.1bp\else -1.1bp\fi\fi}{\bfseries\fontsize{9.5bp}{12bp}\selectfont\color{cnavy}#1}\par\nobreak\vspace{0.2bp}}
\newcommand{\tablehead}[1]{\par\vspace{\iftablelast 8.2bp\else 2.5bp\fi}{\bfseries\fontsize{10.5bp}{13bp}\selectfont\color{cnavy}#1}\par\nobreak}
\newcommand{\bluehead}[1]{\par\vspace{2bp}{\bfseries\fontsize{11bp}{13.5bp}\selectfont\color{cblue}#1}\par\nobreak}
\newcommand{\runin}[1]{\textbf{#1}}
\newcommand{\unnumbered}[1]{\section*{#1}\addcontentsline{toc}{section}{#1}}

\iflatexml\else

\newsavebox{\tocnum}
\renewcommand{\cftsecpresnum}{\global\setbox\tocnum\hbox\bgroup}
\renewcommand{\cftsecaftersnum}{.\egroup}

\renewcommand{\cftsubsecpresnum}{\global\setbox\tocnum\hbox\bgroup}
\renewcommand{\cftsubsecaftersnum}{\egroup}

\makeatletter\renewcommand{\@tocrmarg}{14bp}\renewcommand{\@pnumwidth}{1.5em}\makeatother
\fi

\DeclareCaptionFont{navy}{\color{cnavy}}
\DeclareCaptionFont{capsize}{\fontsize{9bp}{11bp}\selectfont}
\newcommand{\fignote}[1]{\par\vspace{3.7bp}{\fontsize{8bp}{9.6bp}\selectfont\itshape\color{tgray}#1\par}\vspace{-6bp}}
\newsavebox{\panelbox}
\iflatexml
\newcommand{\twopanel}[5]{\begin{quote}\ifx&#4&\else\textbf{#4}\par\fi #5\end{quote}}
\newenvironment{formulabox}[1]{\begin{quote}\textbf{#1}\par\centering}{\end{quote}}
\newcommand{\formula}[1]{\textbf{#1}\par}
\else
\makeatletter
\newcommand{\twopanel}[5]{%
  \if@nobreak\vskip-3.7bp\def\panelbefore{0bp}\else\par\ifdim\lastskip=0bp\def\panelbefore{9.5bp}\else\vskip-\lastskip\def\panelbefore{7bp}\fi\kern-\prevdepth\fi %
  \begin{tcolorbox}[enhanced,breakable,frame hidden,interior hidden,boxrule=0bp,arc=0bp,boxsep=0bp,colupper=body,
    left=257.3bp,right=7.2bp,top=7.8bp,bottom=7.6bp,before skip=\panelbefore,after skip=16.9bp,lines before break=1,
    underlay={\fill[#2]([xshift=249.7bp]frame.south west) rectangle (frame.north east);
              \fill[#1]([yshift=5.5bp]frame.south west) rectangle ([xshift=249.7bp,yshift=-5.5bp]frame.north west);},
    before upper={\RaggedRight\boxsize\color{body}\setlength{\parskip}{3bp}\clubpenalty=0\widowpenalty=0}]%
  \ifx&#4&\else{\bfseries\fontsize{10bp}{12.5bp}\selectfont\color{#3}#4}\par\vspace{-0.5bp}\fi #5
  \end{tcolorbox}\global\panellasttrue}
\makeatother
\newenvironment{formulabox}[1]
  {\par\vspace{1.1bp}\noindent\begin{lrbox}{\panelbox}\begin{minipage}{484.4bp}\setlength{\parskip}{3bp}%
   {\bfseries\fontsize{9.5bp}{12bp}\selectfont\color{cblue}#1}\par\centering}
  {\end{minipage}\end{lrbox}\fboxsep=7.5bp\colorbox{tblue}{\usebox{\panelbox}}\par\vspace{10.4bp}}
\newcommand{\formula}[1]{{\caladea\bfseries\boldmath\fontsize{9.5bp}{12.5bp}\selectfont\color{cnavy}#1\par}}
\fi
\newcommand{\bluepanel}[2]{\twopanel{pblue}{tblue}{cblue}{#1}{#2}}
\newcommand{\greenpanel}[2]{\twopanel{pgreen}{tgreen}{pgreen}{#1}{#2}}
\newcommand{\goldpanel}[2]{\twopanel{pgold}{tgold}{pgold}{#1}{#2}}
\newcommand{\redpanel}[2]{\twopanel{pred}{tred}{pred}{#1}{#2}}

\arrayrulecolor{black}
\newlength{\padtop}\newlength{\padbot}
\newcommand{\tablepad}[2]{\setlength{\padtop}{#1}\setlength{\padbot}{#2}}
\newcommand{\cellpadtop}{\rule{0bp}{\padtop}}
\newcommand{\cellpadbot}{\rule[-\padbot]{0bp}{0bp}}
\iflatexml\else\AfterEndEnvironment{longtable}{\global\tablelasttrue}\fi
\newcolumntype{M}[3]{>{\RaggedRight\arraybackslash\fontsize{#2bp}{#3bp}\selectfont\cellpadtop}m{#1}<{\cellpadbot}}
\newcolumntype{B}[3]{>{\RaggedRight\arraybackslash\fontsize{#2bp}{#3bp}\selectfont\bfseries\color{cnavy}\cellpadtop}m{#1}<{\cellpadbot}}
\newcommand{\thc}[2][8]{\rule[-\dimexpr\padbot+0.4bp\relax]{0bp}{\dimexpr\padtop+1bp\relax}\textcolor{white}{\bfseries\fontsize{#1bp}{10bp}\selectfont #2}}
\newcommand{\headrow}{\rowcolor{thnavy}}
\newcommand{\altrow}{\rowcolor{rowalt}}

\title{Provable Compliance: A Protocol for Runtime Compliance Evidence and Transaction-Specific Decisioning}
\newcommand{\affil}{ChainIT, Utah, United States (\url{https://chainit.com})}%
\author{Jeremy Blackburn\thanks{\affil} \and Tim Hines\thanks{\affil} \and Matthew Koepp\thanks{\affil} \and Russell Lessard\thanks{\affil} \and Jason Patrick\thanks{\affil}}
\date{September 2026}

\iflatexml\else
\renewenvironment{abstract}{\section*{Abstract}}{\par}
\fi

\begin{document}

\iflatexml
\maketitle
\else
\thispagestyle{empty}
\setlength{\TPHorizModule}{1bp}\setlength{\TPVertModule}{1bp}
\newcommand{\coverline}[3]{%
  \begin{textblock*}{499.4bp}(56.3bp,#1bp)\centering #2 #3\par\end{textblock*}}
\coverline{59.7}{\bfseries\fontsize{30bp}{36bp}\selectfont\color{cnavy}}{Provable Compliance}
\coverline{106.5}{\fontsize{12.5bp}{17bp}\selectfont\color{cgray}}{A Protocol for Runtime Compliance Evidence and Transaction-Specific Decisioning}
\coverline{199.5}{\bfseries\fontsize{15bp}{18bp}\selectfont\color{cgray}}{ChainIT}
\coverline{230.0}{\bfseries\fontsize{12bp}{14.5bp}\selectfont\color{cblue}}{\url{https://chainit.com}}
\coverline{315.5}{\bfseries\fontsize{11bp}{13.5bp}\selectfont\color{cnavy}}{ChainIT Compliance Protocol — CIT-PC-WP-4.0}
\coverline{334.7}{\itshape\fontsize{9.5bp}{11.5bp}\selectfont\color{cgray}}{Architecture and Source Basis}
\coverline{406.0}{\fontsize{10.5bp}{12.5bp}\selectfont\color{hgray}}{September 2026}
\coverline{424.0}{\itshape\fontsize{8bp}{9.6bp}\selectfont\color{hgray}}{Companion document: ChainIT \href{https://chainit.com/insights/chainit-launches-provable-authority-as-u-s-generative-ai-enabled-fraud-losses-are-projected-to-reach-40-billion-by-2027/}{Provable Authority} (CIT-PA-WP-7.0).}
\coverline{493.0}{\bfseries\fontsize{10.5bp}{12.5bp}\selectfont\color{body}}{Authors}
\coverline{512.1}{\fontsize{9.5bp}{13bp}\selectfont\color{body}}{\href{https://chainit.com/about-us/}{Jeremy Blackburn}, Chief Executive Officer\\\href{https://chainit.com/about-us/}{Tim Hines}, Chief Product Officer\\\href{https://chainit.com/about-us/}{Matthew Koepp}, Chief Technology Officer\\\href{https://chainit.com/about-us/}{Russell Lessard}, Chief Compliance Officer\\\href{https://chainit.com/about-us/}{Jason Patrick}, Chief Commercial Officer}
\coverline{693.6}{\fontsize{8bp}{9.5bp}\selectfont\color{cgray}}{This paper describes the ChainIT \href{https://chainit.com/white-papers/}{Provable Compliance} architecture. It does not constitute legal, regulatory, banking,\\accounting, security-certification, or patentability advice, and it does not replace a regulated institution’s own AML/CFT\\or sanctions program. Deployment configurations, provider coverage, and service levels vary by implementation.}
\null\clearpage
\fi

\iflatexml\else
\begin{center}
{\bfseries\fontsize{11bp}{13bp}\selectfont\color{black}\textcopyright{} 2026, Scientia Potentia Est.\ II, LLC. All rights reserved.\par}
\vspace{2bp}
{\fontsize{8bp}{9.6bp}\selectfont\color{cgray}ChainIT is a registered trademark of Scientia Potentia Est.\ II, LLC. Distributed by arXiv under the arXiv.org perpetual, non-exclusive license.\par}
\end{center}
\color{body}
\fi

\begin{abstract}
Regulated institutions already perform onboarding, screening, monitoring, and review, but compliance evidence ages, and whether it supports a decision depends on the institution, product, subject, counterparty, transaction, jurisdiction, risk tier, and purpose. A relying party therefore often cannot obtain compact, attributable proof of which evidence, policy, authority, and transaction context were current when a decision was made, a gap that grows as authority is delegated to software agents and scheduled workflows. We describe ChainIT \href{https://chainit.com/white-papers/}{Provable Compliance}, the compliance-side layer of the ChainIT \href{https://chainit.com/insights/chainit-launches-provable-authority-as-u-s-generative-ai-enabled-fraud-losses-are-projected-to-reach-40-billion-by-2027/}{Authority Protocol}. Each sourced or institution-generated fact is represented as a separate Evidence Assertion \href{https://chainit.com/product/data-tokenization/}{VDT} with provenance, freshness, confidence, permitted purpose, and privacy classification; a versioned Compliance Policy Profile selects the applicable rules; a Business Rules Engine returns deterministic, reason-coded predicate results; and the organization's configured Workflow Engine composes them into a signed, time-bounded Compliance Decision VDT for one reliance event. The decision contributes a single predicate, Compliance Decision Permits, to session-eligibility and transaction-policy decisions. It never independently authorizes payment: only a final Allow together with an exact transaction digest, atomic capacity reservation, and a single-use execution credential can commit value. The paper specifies point-of-reliance freshness, scoped propagation of evidence changes, decision states, evidence domains (AML, sanctions, PEP, background, identity, and licensing), the U.S. and FATF regulatory context as of August 2026, privacy and disclosure controls, and residual risks.
\end{abstract}

\textbf{Keywords:} compliance evidence; runtime verification; AML/CFT; sanctions screening; verifiable credentials; policy-based decisioning; agentic commerce; stablecoins.

{\fontsize{9.5bp}{12.8bp}\selectfont\color{cgray}This paper was first released as a ChainIT white paper (CIT-PC-WP-4.0). The companion paper, \emph{Provable Authority} \cite{ref23}, and other ChainIT white papers are listed at \url{https://chainit.com/white-papers/}.\par}
\clearpage

\iflatexml\else
\tableofcontents
\clearpage
\fi

\unnumbered{Definitions and Document Precedence}\label{sec:definitions}

\bluepanel{Authority definitions control}{For all terms, artifacts, subject classifications, and process steps defined in ChainIT \href{https://chainit.com/insights/chainit-launches-provable-authority-as-u-s-generative-ai-enabled-fraud-losses-are-projected-to-reach-40-billion-by-2027/}{Provable Authority}, CIT-PA-WP-7.0 controls. This paper adopts those definitions and flows by reference and defines only the compliance-side evidence, policy, and decision architecture. Any abbreviated description in this paper is explanatory and does not modify the Authority Protocol. \cite{ref23}}

The shared \href{https://chainit.com/insights/chainit-launches-provable-authority-as-u-s-generative-ai-enabled-fraud-losses-are-projected-to-reach-40-billion-by-2027/}{Authority Protocol} terms below are reproduced or closely restated from CIT-PA-WP-7.0 so that the two papers can be read together without creating a second or conflicting authority model. Compliance-specific terms are then defined for this paper. Definitions are provided for reader clarity and do not themselves constitute legal, regulatory, product, or security specifications.

\setlength{\LTleft}{-51.0bp}
\tablepad{12.3bp}{7.0bp}
\begin{longtable}{|B{238.9bp}{7.5}{10.15}|M{340.5bp}{7.5}{10.15}|}
\hline
\headrow\thc[7.5]{Term} & \thc[7.5]{Definition}\\\hline
\endfirsthead
\hline
\headrow\thc[7.5]{Term} & \thc[7.5]{Definition}\\\hline
\endhead
Agent Authority \href{https://chainit.com/product/data-tokenization/}{VDT} & The append-only record of a grant of delegated authority from an authorized human or organizational principal to a registered agent, and the agent-specialized form of the Delegated Authority VDT. Each entry immutably records the granting principal, the governing ARP and its version, the registered agent, permitted purpose, counterparties, transaction classes, rails, per-transaction and aggregate value limits, effective period, step-up thresholds, redelegation rule, and revocation conditions. An agent holds no authority of its own. Present validity is resolved at the reliance event against current ARP status and the applicable freshness bound. It is not a bearer credential.\\\hline
\altrow Agent Registration \href{https://chainit.com/product/data-tokenization/}{VDT} & A versioned record identifying an autonomous workload’s organization, responsible party or role, tenant, deployment, software or workload identity, permitted tools and APIs, risk class, attestation method, registered public key, status reference, and version history. It establishes what the workload is, not what it may do.\\\hline
Agent Subject Profile & The identity and containment controls that exist because the acting subject is an autonomous workload rather than a person: the Agent Registration \href{https://chainit.com/product/data-tokenization/}{VDT}, the attested workload key and ephemeral session-key hierarchy, and the sovereignty boundary. The profile adds no authority concept of its own; an agent exercises delegated authority under the ARP in the same way a human delegate does, evidenced by an Agent Authority VDT.\\\hline
\altrow Authority Policy Profile (APP) & The versioned rule set selected for a given reliance event, which may vary by organization, product, role, subject class, counterparty, transaction type, value, payment rail, jurisdiction, risk tier, and intended purpose. It is the authority-side counterpart to the Compliance Policy Profile and is versioned so that a later reviewer can determine exactly which rule set produced an authority decision.\\\hline
\href{https://chainit.com/insights/chainit-launches-provable-authority-as-u-s-generative-ai-enabled-fraud-losses-are-projected-to-reach-40-billion-by-2027/}{Authority Protocol} & The general ChainIT protocol for resolving whether an identified subject may perform a specific proposed action at the moment of reliance. It is subject-neutral and rail-neutral: the same identity, authority, evidence, policy, digest, reservation, execution, and outcome-evidence controls govern individual-, group-, organization-, and agent-initiated actions, including organization-controlled scheduled-workflow initiations.\\\hline
\altrow Authority Resolution \href{https://www.pactvera.com/}{Pactvera} (ARP) & The canonical, versioned organizational authority record: who may act, in what role, for what purpose, within what limits, during what effective period, under what quorum or separation-of-duties requirements, and subject to what amendment or revocation.\\\hline
Business Rules Engine (BRE) & The deterministic predicate-evaluation engine. It evaluates configured conditions against subject identity and assurance, ARP authority, registration and attestation where the subject is a workload, evidence, status freshness, jurisdiction, scope, session risk, and—for a transaction—the exact transaction, current capacity, required approvals, and rail-specific constraints. It returns explicit, reason-coded true/false predicate results. The BRE does not itself perform risk scoring or assign a disposition; it runs within the ChainIT Workflow Engine.\\\hline
\altrow Canonical transaction digest & A deterministic hash of a canonical transaction object—including payer, source, payee, destination, amount, currency or asset, rail, fees, nonce, validity window, and rail-specific parameters—computed under a fixed encoding rule so that any material change produces a different digest.\\\hline
\href{https://chainit.com/product/identity-verification/}{ChainIT ID} & The verified identity anchor for an individual, established through authoritative-source verification, biometric and liveness controls, and privacy-preserving, portable identity storage.\\\hline
\altrow \href{https://chainit.com/product/business-verification/}{ChainIT Org ID} & The legal-entity identity anchor for an organization, which may reference entity registration, ownership, officers, governance, and authorized representatives.\\\hline
\href{https://chainit.com/product/digital-pay/}{ChainIT Pay} & The ChainIT payment-execution platform intended to serve as the EAC enforcement boundary where the applicable \href{https://chainit.com/insights/chainit-launches-provable-authority-as-u-s-generative-ai-enabled-fraud-losses-are-projected-to-reach-40-billion-by-2027/}{Authority Protocol} profile is deployed; a compatible partner rail adapter may serve this function in another implementation.\\\hline
\altrow Compliance Decision Permits & The compliance-side predicate consumed by the \href{https://chainit.com/insights/chainit-launches-provable-authority-as-u-s-generative-ai-enabled-fraud-losses-are-projected-to-reach-40-billion-by-2027/}{Authority Protocol}. It is true only when the applicable Compliance Decision \href{https://chainit.com/product/data-tokenization/}{VDT} has a final Allow disposition for the stated reliance event after every required compliance control has been completed and the decision has been re-evaluated. It does not establish authority, session eligibility, transaction authority, capacity, or payment authority.\\\hline
Compliance Decision \href{https://chainit.com/product/data-tokenization/}{VDT} & The signed, time-bounded, purpose-specific record produced by an organization’s configuration of the ChainIT Workflow Engine, recording whether present evidence and policy permit a proposed exercise of authority. It gates present exercise; it does not amend, suspend, or erase the governing ARP. An Allow disposition means compliance-side continuation only and never independently authorizes payment.\\\hline
\altrow Compliance Policy Profile (CPP) & The versioned rule set selecting the compliance evidence, freshness bounds, predicates, review conditions, disclosure rules, and other compliance-side requirements applicable to a reliance event. It may vary by institution, regulated entity, product, role, subject, counterparty, transaction type, value, rail, payer and payee jurisdiction, customer category, risk tier, and intended purpose.\\\hline
Delegated Authority \href{https://chainit.com/product/data-tokenization/}{VDT} & A portable artifact that references the governing ARP and exposes only the scope of authority needed by a given delegate, agent, or relying party without replacing or superseding the ARP itself. Its agent-specialized form is the Agent Authority VDT.\\\hline
\altrow Disposition & The explicit outcome produced by an organization’s configuration of the ChainIT Workflow Engine from BRE predicate results: Allow, Allow with Controls, Step-Up, Review, Hold, Reject, Prohibit, or Stale/Unknown. The organization configures the disposition set to match its own policies and regulatory obligations. Disposition and risk tier are evaluated independently.\\\hline
Evidence Assertion \href{https://chainit.com/product/data-tokenization/}{VDT} & A record of a single sourced fact or institution-generated state, including its subject, source class, source, source version, retrieval and effective time, expiration or review time, jurisdiction, match or confidence state, permitted purpose, and privacy classification. Evidence Assertion VDTs are never aggregated into a universal “compliant” status.\\\hline
\altrow Execution Authorization Credential (EAC) & A single-use, short-lived credential authorizing one exact transaction for one audience, bound to the acting session or approved subject, transaction digest, reservation, status epoch, and expiration, and consumed atomically at commit.\\\hline
Execution Authorization Lifecycle \href{https://chainit.com/product/data-tokenization/}{VDT} & A permanent, append-only evidence record of EAC issuance and terminal handling—consumed, expired, canceled, or rejected. The EAC itself is not stored as a reusable ledger object, and earlier lifecycle events are never modified or overwritten.\\\hline
\altrow \href{https://www.pactvera.com/}{Pactvera} & The ChainIT transaction-logic and append-only evidence framework linking \href{https://chainit.com/product/data-tokenization/}{Validated Data Tokens} across a transaction’s lifecycle.\\\hline
\href{https://chainit.com/white-papers/}{Provable Compliance} & The property of a system in which compliance-relevant evidence, rule selection, predicate evaluation, resulting disposition, freshness, and disclosure treatment are attributable, time-bounded, purpose-specific, and independently verifiable. Provable Compliance supplies the compliance-side decision consumed by the \href{https://chainit.com/insights/chainit-launches-provable-authority-as-u-s-generative-ai-enabled-fraud-losses-are-projected-to-reach-40-billion-by-2027/}{Authority Protocol}; it does not itself confer authority.\\\hline
\altrow Reliance event & The moment at which a relying party depends on identity, authority, evidence, or policy state—session issuance or renewal, a transaction proposal or payment initiation, a human step-up approval, a scheduled review, or a source-change or material-risk event. Freshness and policy are evaluated at the reliance event rather than inherited from onboarding.\\\hline
Reservation & The atomic hold of transaction capacity against a budget or limit, performed before EAC issuance so concurrent requests cannot commit against the same capacity twice.\\\hline
\altrow Risk tier & A classification—Tier 1–Low, Tier 2–Standard, or Tier 3–High—assigned to a transaction based on factors such as payee history, value, reversibility, and behavioral or device anomalies. It determines minimum verification rigor. The tier is assigned by the organization’s configuration of the ChainIT Workflow Engine, not by the BRE, and it does not itself determine the disposition.\\\hline
Session Authorization Credential (SAC) & A short-lived, sender-constrained credential that permits proposal activity, evidence retrieval, and defined API activity, but which rail adapters must reject as payment authority. A session credential is never payment authority, whether the session belongs to a workload or a signed-in person.\\\hline
\altrow Session Eligibility Decision \href{https://chainit.com/product/data-tokenization/}{VDT} & A session-context decision recording whether the authenticated subject may open or continue a session that enables consequential action, referencing identity assurance, the ARP, evidence, status, jurisdiction, scope, and policy version. Where the subject is a workload, it additionally references the Agent Registration VDT, Agent Authority VDT, attested deployment, and applicable tool scope. It is not transaction-specific and cannot authorize payment.\\\hline
Sovereignty boundary & The architectural line beyond which an AI model or autonomous agent may reason, gather evidence, and propose an action, but cannot obtain unrestricted access to payment-account credentials, custody keys, or final execution authority; that authority remains with the deterministic control plane.\\\hline
\altrow Status epoch / status freshness & The signed timestamp or snapshot boundary used to determine whether cached non-revocation or evidence status is still acceptable for a given risk tier versus requiring a live status or source interaction.\\\hline
Subject of authority & The party whose authority is being resolved: a verified individual, a group acting under a quorum or separation-of-duties rule, an organization acting through an authorized representative, or a permitted software agent. A scheduled workflow is not a fifth subject class; it is an organization-controlled initiation mechanism acting for an identified organization under the governing ARP.\\\hline
\altrow Supplemental \href{https://chainit.com/}{Valitorum} & A permanent sealed evidence package for a later post-terminal exception—such as a return, reversal, recall, chargeback, or dispute—that references the original Valitorum without altering it.\\\hline
Transaction Policy Decision \href{https://chainit.com/product/data-tokenization/}{VDT} & A decision bound to one canonical transaction digest, recording risk tier, disposition, policy version, evidence references, ARP and subject references, required approvals or controls, expiration, and reason codes. It is newly derived per transaction and is not reusable as a standing status.\\\hline
\altrow \href{https://chainit.com/product/data-tokenization/}{Validated Data Token} (VDT) & A cryptographically verifiable digital record representing validated data concerning a person, organization, asset, service, event, decision, or transaction. A change is represented by a new linked VDT rather than alteration of a prior immutable record.\\\hline
\href{https://chainit.com/}{Valitorum} & The final sealed evidence record generated for a completed or otherwise terminal transaction lifecycle under a defined completeness rule. A sealed Valitorum is never altered; later exceptions are captured in a linked Supplemental Valitorum.\\\hline
\altrow \href{https://chainit.com/}{Valitorum} Service & The ChainIT service that creates, verifies, retrieves, or presents a Valitorum and, where applicable, a Supplemental Valitorum.\\\hline
Workflow Engine & The organization-configurable orchestration layer in which BRE predicate results, risk-tier assignment, required approvals, review states, and other configured conditions are composed into a disposition and downstream action. The organization is responsible for the policy logic it configures and for its use of the resulting disposition, subject to applicable law and contractual allocation among ChainIT, the organization, payment providers, evidence providers, and other participating parties.\\\hline
\end{longtable}
\clearpage

\unnumbered{Executive Summary}\label{sec:summary}

\href{https://chainit.com/insights/chainit-launches-provable-authority-as-u-s-generative-ai-enabled-fraud-losses-are-projected-to-reach-40-billion-by-2027/}{Provable Authority} asks whether an identified individual, group, organization, or permitted software agent is authorized to perform a particular action. \href{https://chainit.com/white-papers/}{Provable Compliance} addresses a different question: whether the evidence and policy conditions required for the present action are sufficiently current, complete, and favorable for that authority to be exercised now. The final definitions and execution flows for the shared Authority Protocol are those in CIT-PA-WP-7.0; this paper supplies the compliance-side evidence and decision architecture consumed by those flows. \cite{ref23}

The distinction matters because compliance is neither a permanent badge nor a universal statement that a person or organization is “in good standing to do anything at all.” AML monitoring, sanctions, politically exposed person status, criminal-history information, government-credential state, licensing, KYC/KYB, and payment-transparency obligations operate differently. Some are authoritative facts, some are qualified third-party indicators, and some are institution-generated judgments. Their consequences depend on the institution, product, role, subject class, transaction, counterparty, jurisdiction, risk tier, and intended purpose.

The ChainIT model therefore uses modular Evidence Assertion \href{https://chainit.com/product/data-tokenization/}{VDTs} and a reliance-event-specific Compliance Decision VDT. \href{https://chainit.com/product/identity-verification/}{ChainIT ID} and \href{https://chainit.com/product/business-verification/}{ChainIT Org ID} establish verified subjects. The ARP remains the canonical organizational authority record, and a Delegated Authority VDT—or its agent-specialized Agent Authority VDT—exposes a bounded grant without replacing the ARP. Current Evidence Assertion VDTs preserve source, provenance, retrieval and effective time, confidence, permitted purpose, and privacy classification. A versioned CPP selects the applicable conditions. The BRE returns explicit, deterministic, reason-coded true/false predicate results. The organization’s configuration of the ChainIT Workflow Engine composes those results into a signed, time-bounded Compliance Decision VDT.

A favorable Compliance Decision \href{https://chainit.com/product/data-tokenization/}{VDT} does not establish session eligibility or authorize a transaction. It contributes the Compliance Decision Permits predicate to the final \href{https://chainit.com/insights/chainit-launches-provable-authority-as-u-s-generative-ai-enabled-fraud-losses-are-projected-to-reach-40-billion-by-2027/}{Authority Protocol}. For a session, that predicate is evaluated with subject authentication, ARP or delegated authority, the APP, status, scope, and—where applicable—the Agent Subject Profile to produce a Session Eligibility Decision VDT and, only after a final Allow, a SAC. For a transaction, it is evaluated with the canonical transaction digest, authority, APP, risk tier, current capacity, required approvals, and rail constraints to produce the Transaction Policy Decision VDT. Controls and step-up must be completed and the transaction re-evaluated to final Allow before capacity reservation and EAC issuance. \cite{ref23}

Runtime verification means that ChainIT evaluates evidence freshness and policy at the moment of reliance. It does not mean every external source is queried for every session. Each evidence domain has a source-specific and risk-specific freshness requirement. ChainIT may reuse a signed Evidence Assertion \href{https://chainit.com/product/data-tokenization/}{VDT} while it remains fresh enough for the present context, perform a live or event-driven refresh when required, and produce Review, Hold, or Stale/Unknown when a source is unavailable, stale, or ambiguous. A change propagates only to the subject, activity, product, rail, counterparty, jurisdiction, or authority relationship that actually depends on it unless law or policy requires broader blocking.

The resulting compliance decision and its evidence references become part of the same append-only lifecycle used by \href{https://chainit.com/insights/chainit-launches-provable-authority-as-u-s-generative-ai-enabled-fraud-losses-are-projected-to-reach-40-billion-by-2027/}{Provable Authority}. The operational EAC is not stored as a reusable ledger object; Execution Authorization Lifecycle \href{https://chainit.com/product/data-tokenization/}{VDTs} record issuance and terminal handling. Each later transaction state is a new event. The \href{https://chainit.com/}{Valitorum} seals the completed or otherwise terminal evidence graph, and a later return, reversal, recall, chargeback, or dispute may be sealed in a Supplemental Valitorum that references—but never modifies—the original.

\greenpanel{Core proposition}{Provable Compliance does not ask whether a subject is universally “compliant.” It produces a signed, time-bounded, purpose-specific decision stating whether identified evidence and an identified policy permit compliance-side continuation for one reliance event. The final \href{https://chainit.com/insights/chainit-launches-provable-authority-as-u-s-generative-ai-enabled-fraud-losses-are-projected-to-reach-40-billion-by-2027/}{Authority Protocol} still decides whether the subject may open a consequential session or commit an exact transaction.}

\section{The Problem: Compliance Evidence Ages and Context Changes}\label{sec:problem}

The concern ChainIT addresses is not that regulated institutions perform no ongoing monitoring. Banks, payment companies, money-services businesses, and other regulated organizations already maintain onboarding controls, transaction monitoring, sanctions screening, case management, customer-risk profiles, periodic or event-driven reviews, and escalation processes. U.S. customer-due-diligence rules expressly require ongoing monitoring and risk-based updating of customer information; they do not require every data element to be refreshed on one universal schedule. \cite{ref1}

The operational gap is that relevant facts and judgments remain fragmented. Identity evidence may be in one system, organizational authority in another, sanctions results in a third, AML alerts in a fourth, and payment approval in a bank portal or ERP application. Even where each control is competently operated, a relying party may not be able to obtain compact, cryptographically attributable proof showing which evidence, policy, authority, and transaction context were current at the moment a decision was made.

The gap becomes more consequential as authority is delegated to software agents and scheduled workflows. A human may informally notice an unfamiliar payee, a changed bank account, an expired license, or an unusual pattern. An automated workflow executes the tools, data, and permissions it is given. ChainIT therefore places the same deterministic compliance and authority boundary in front of an agent proposal, a human-initiated wire, a group-approved ACH file, an organization-controlled scheduled payment, and an on-chain transfer. The subject authentication differs; the governing evidence, authority, policy, and execution controls do not.

The objective is not to replace an institution’s AML/CFT, sanctions, licensing, privacy, or other compliance program. It is to make the institution’s evidence, policy selection, predicate evaluation, disposition, authority context, and reliance event machine-verifiable at the moment of use and auditable after the event.

\greenpanel{}{\vspace{-3bp}\begin{itemize}\boxsize
  \item Ongoing monitoring is not the same as a permanent status; evidence that was accurate at onboarding does not remain accurate by default.
  \item A favorable historical screening result does not establish that a subject is currently clear; sanctions, PEP, and entity status can change without notice.
  \item Compliance evidence gathered for one purpose or product does not automatically justify reliance for a different purpose, counterparty, or jurisdiction.
  \item A policy exception approved for one reliance event does not carry forward to the next unless the applicable CPP says it does.
  \item An unavailable or ambiguous evidence source is not the same as a favorable result; it requires a controlled Review, Hold, or Stale/Unknown disposition.
\end{itemize}}

\subsection*{Capability Status}

This paper describes present-tense architecture to keep the control model legible. Not every artifact below is deployed in production in every implementation. Each principal component below is classified on two independent axes: whether it is built and deployed today, and what patent or documentary support currently exists for it. These are different questions — a capability may be deployed while representing potential new patent matter, or described in a filed application without yet being built. The matrix below is the single authoritative classification for this component set; the companion Pilot Measurement \& Validation Plan (CIT-PC-PILOT-4.0) uses the identical classification, and the companion Architecture Diagrams \& Flows (CIT-PC-DIAGRAMS-4.0) intentionally omits an implementation-status legend to avoid a second, competing classification.

\goldpanel{Pending formal sign-off}{This matrix reflects the current working classification. Before external distribution, Product, Engineering, and Patent Counsel should formally confirm each row and any change should be reproduced identically across all three companion documents.}

\minorhead{Implementation status}
{\smallsize Production: deployed and in current use. Development: being built or validated, not yet in general production use. Proposed: architecture is defined but not yet built. External dependency: supplied by a third party outside ChainIT\textquotesingle{}s direct control.\par}

\minorhead{IP / documentary support}
{\smallsize Issued patent: covered by an issued ChainIT patent. Filed application: described in a filed, not-yet-issued application. Potential new matter: not yet filed; a candidate for new filing. Counsel review: filing posture not yet determined. Not applicable: a third-party capability outside ChainIT\textquotesingle{}s patent estate.\par}

\vspace{19.8bp}%
\setlength{\LTleft}{13.7bp}
\tablepad{11.3bp}{4.8bp}
\begin{longtable}{|B{199.2bp}{7.5}{9}|M{124.9bp}{7.5}{9}|M{114.9bp}{7.5}{9}|}
\hline
\headrow\thc{Component} & \thc{Implementation status} & \thc{IP / documentary support}\\\hline
\endfirsthead
\hline
\headrow\thc{Component} & \thc{Implementation status} & \thc{IP / documentary support}\\\hline
\endhead
\href{https://chainit.com/product/identity-verification/}{ChainIT ID} / \href{https://chainit.com/product/business-verification/}{ChainIT Org ID} / ARP & Production & Issued patent\\\hline
\altrow \href{https://www.pactvera.com/}{Pactvera} / \href{https://chainit.com/}{Valitorum} & Production & Issued patent\\\hline
Business Rules Engine (compliance predicates) & Production & Issued patent\\\hline
\altrow Evidence Assertion \href{https://chainit.com/product/data-tokenization/}{VDT} & Development & Filed application\\\hline
Compliance Policy Profile (CPP) & Development & Filed application\\\hline
\altrow Status epoch / freshness model & Development & Filed application\\\hline
ChainIT Workflow Engine (compliance disposition) & Development & Potential new matter\\\hline
\altrow Compliance Decision \href{https://chainit.com/product/data-tokenization/}{VDT} (issuer / signature) & Development & Potential new matter\\\hline
Compliance Decision Permits interface & Proposed & Potential new matter\\\hline
\altrow Disclosure-tier engine & Development & Potential new matter\\\hline
Manual-review and override architecture & Proposed & Counsel review\\\hline
\altrow Supplemental \href{https://chainit.com/}{Valitorum} & Development & Filed application\\\hline
Evidence source integrations (sanctions, PEP, registries, bureaus) & External dependency & Not applicable\\\hline
\altrow Partner rail-adapter enforcement & External dependency & Not applicable\\\hline
\end{longtable}

\section{The ChainIT Model: Evidence, Policy, Authority, and Decision}\label{sec:model}

\href{https://chainit.com/white-papers/}{Provable Compliance} is one layer inside the final ChainIT \href{https://chainit.com/insights/chainit-launches-provable-authority-as-u-s-generative-ai-enabled-fraud-losses-are-projected-to-reach-40-billion-by-2027/}{Authority Protocol}. It collects or references modular evidence, applies a versioned compliance policy, evaluates deterministic predicates, and records a purpose-specific disposition. The Authority Protocol then consumes that compliance result together with identity, authority, session or transaction context, risk tier, approvals, capacity, and rail controls. Compliance and authority are adjacent but independent: authority can exist while its present exercise is held, and favorable compliance evidence cannot create authority that the ARP does not grant. \cite{ref23}

\subsection{Subject neutrality and verified identity foundation}\label{sec:subjects}

The \href{https://chainit.com/insights/chainit-launches-provable-authority-as-u-s-generative-ai-enabled-fraud-losses-are-projected-to-reach-40-billion-by-2027/}{Authority Protocol} recognizes four subject classes: a verified individual; a group acting under quorum or separation-of-duties rules; an organization acting through an authorized representative; and a permitted software agent. A scheduled workflow is not a fifth subject class. It is an organization-controlled initiation mechanism acting for an identified organization under the governing ARP. \cite{ref23}

\href{https://chainit.com/product/identity-verification/}{ChainIT ID} is the verified identity anchor for an individual. \href{https://chainit.com/product/business-verification/}{ChainIT Org ID} is the legal-entity identity anchor for an organization. The identity materials support authoritative-source identity verification, biometric authentication and liveness, selective disclosure, lifecycle updates, and portable or offline verification without requiring a new government query every time identity is presented. \cite{ref17} Identity and authority remain distinct: ChainIT Org ID establishes the entity; the ARP establishes who or what may act for it.

\subsection{ARP, Delegated Authority, and Agent Authority}\label{sec:arp}

The ARP is the canonical, versioned organizational authority record. It identifies who may act, in what role, for what purpose, within what limits, during what effective period, under what quorum or separation-of-duties requirements, and subject to what amendment or revocation. \cite{ref18,ref19,ref23} A Delegated Authority \href{https://chainit.com/product/data-tokenization/}{VDT} references the ARP and exposes only the bounded scope needed by a delegate or relying party. Where the delegate is a registered agent, the Agent Authority VDT is the agent-specialized form of that portable authority artifact.

An agent has no authority of its own. Its Agent Authority \href{https://chainit.com/product/data-tokenization/}{VDT} records the granting principal, governing ARP and version, registered agent, purpose, counterparties, transaction classes, rails, value limits, effective period, step-up thresholds, redelegation rule, and revocation conditions. The grant is append-only; current validity is resolved at the reliance event against the ARP and applicable freshness. A compliance disposition may condition or prevent present exercise, but it does not amend, suspend, delete, or retroactively erase the ARP or the grant record.

\subsection{Evidence Assertion VDTs}\label{sec:evidence}

An Evidence Assertion \href{https://chainit.com/product/data-tokenization/}{VDT} represents one sourced observation or institution-generated state, not the final compliance conclusion. It identifies the subject, source class, source, source version, retrieval time, effective time, expiration or review time, jurisdiction, match or confidence state, permitted purpose, privacy classification, and integrity reference. Separate VDTs preserve the differences among identity-document status, sanctions data, PEP information, professional licensing, KYB evidence, AML monitoring state, case status, and transaction-specific counterparty facts.

\setlength{\LTleft}{-15.5bp}
\tablepad{12.8bp}{7.2bp}
\begin{longtable}{|B{155.6bp}{8}{10.85}|M{172.7bp}{8}{10.85}|M{169.2bp}{8}{10.85}|}
\hline
\headrow\thc{Source class} & \thc{Examples} & \thc{Required treatment}\\\hline
\endfirsthead
\hline
\headrow\thc{Source class} & \thc{Examples} & \thc{Required treatment}\\\hline
\endhead
Authoritative-source evidence & Government or statutory registries; issuing-jurisdiction identity data; professional-license records; official sanctions lists. & Record the exact source, participating jurisdiction or program, fields returned, version, time, and limits of what the source proves.\\\hline
\altrow Qualified third-party evidence & PEP data; sanctions and ownership enrichment; criminal-history or background reports; adverse-media services; certifications and specialist providers. & Record provider, permissible purpose, methodology or match state, confidence, dispute path, and contractual or legal restrictions.\\\hline
Institution-generated evidence & Transaction monitoring; customer and relationship risk; source-of-funds analysis; internal alerts; case-management state; analyst findings. & Identify the responsible institution, model or rule version, analyst or case state, confidentiality class, and permitted disclosure.\\\hline
\end{longtable}

\subsection{Compliance Policy Profile}\label{sec:cpp}

The CPP selects the compliance-side rules for a reliance event. It may vary by institution, regulated entity, product, role, subject class, counterparty, transaction type, value, payment rail, payer and payee jurisdictions, customer category, risk tier, and intended purpose. It identifies required evidence, acceptable source classes, maximum evidence age, live-query conditions, predicate logic, review requirements, disposition mapping, disclosure tier, retention, and degraded-mode treatment. The profile is versioned so a later reviewer can determine exactly which rule set produced the result. It is the compliance-side counterpart to the APP; neither profile creates the authority grant recorded by the ARP. \cite{ref23}

\subsection{BRE, Workflow Engine, and Compliance Decision VDT}\label{sec:bre}

The BRE returns explicit, deterministic, reason-coded true/false predicate results. It does not itself perform risk scoring or assign the disposition. The organization’s configuration of the ChainIT Workflow Engine selects the risk tier where applicable, composes the predicate results, applies quorum or review logic, and creates the disposition. The signed, time-bounded result is the Compliance Decision \href{https://chainit.com/product/data-tokenization/}{VDT}.

The Compliance Decision \href{https://chainit.com/product/data-tokenization/}{VDT} identifies the subject and reliance event; CPP and version; Evidence Assertion VDT references; evidence ages or status epochs; BRE predicate results; organization-defined disposition; required controls; reason codes; decision time; expiration or reevaluation time; affected scope; review case where applicable; and permitted disclosure level. It contains signed results and controlled references rather than raw source records that a relying party does not need.

Each Compliance Decision \href{https://chainit.com/product/data-tokenization/}{VDT} identifies the issuing organization or authorized decision service, the applicable tenant and regulated entity, the signature key or verification method, the relying-party or audience restriction, the decision identifier, the reliance-event identifier, the subject and transaction or session reference, the CPP version, evidence epoch, issue time, expiration, status reference, disclosure tier, and reason codes. This is distinct from the Evidence Assertion VDTs it references, which may themselves be a source-signed assertion, a ChainIT-created attestation recording an authenticated source response, or an institution-generated evidence assertion; a source or registry does not itself sign a ChainIT VDT.

\goldpanel{Responsibility and allocation}{The organization is responsible for the policy logic it configures and for its use of the resulting disposition, subject to applicable law and contractual allocation among ChainIT, the organization, payment providers, evidence providers, and other participating parties. The white paper does not independently allocate all legal liability.}

\subsection{Compliance as an input to the final Authority Protocol}\label{sec:input}

The Compliance Decision \href{https://chainit.com/product/data-tokenization/}{VDT} is not the final session or transaction decision. An Allow disposition means only that the compliance-side predicates permit continuation. A subject may still lack authority, exceed scope, fail authentication, lack required approval or capacity, present a modified transaction digest, or lack a valid EAC. Only the Session Eligibility Decision VDT or Transaction Policy Decision VDT provides the broader \href{https://chainit.com/insights/chainit-launches-provable-authority-as-u-s-generative-ai-enabled-fraud-losses-are-projected-to-reach-40-billion-by-2027/}{Authority Protocol} disposition for the applicable reliance event. A Compliance Decision VDT never independently authorizes payment. \cite{ref23}

\setlength{\LTleft}{-35.8bp}
\tablepad{12.3bp}{7.0bp}
\begin{longtable}{|B{155.7bp}{7.5}{10.15}|M{226.7bp}{7.5}{10.15}|M{155.6bp}{7.5}{10.15}|}
\hline
\headrow\thc[7.5]{Decision artifact} & \thc[7.5]{Decision scope} & \thc[7.5]{Critical limitation}\\\hline
\endfirsthead
\hline
\headrow\thc[7.5]{Decision artifact} & \thc[7.5]{Decision scope} & \thc[7.5]{Critical limitation}\\\hline
\endhead
Compliance Decision \href{https://chainit.com/product/data-tokenization/}{VDT} & Whether current compliance evidence and the CPP permit, condition, hold, review, reject, or prohibit compliance-side continuation. & Does not establish authority, session eligibility, transaction authority, capacity, or payment authority.\\\hline
\altrow Session Eligibility Decision \href{https://chainit.com/product/data-tokenization/}{VDT} & Whether the authenticated subject may open or continue a consequential session under current identity, authority, evidence, status, scope, and policy. & Not transaction-specific and cannot authorize value movement.\\\hline
Transaction Policy Decision \href{https://chainit.com/product/data-tokenization/}{VDT} & Whether the complete identity, authority, compliance, exact transaction, risk, capacity, approval, and rail conditions support the proposed action. & Only final Allow, after controls and step-up are completed and revalidated, may support reservation and EAC issuance.\\\hline
\end{longtable}

\begin{figure}[H]
  \centering
  \includegraphics[width=421.16bp]{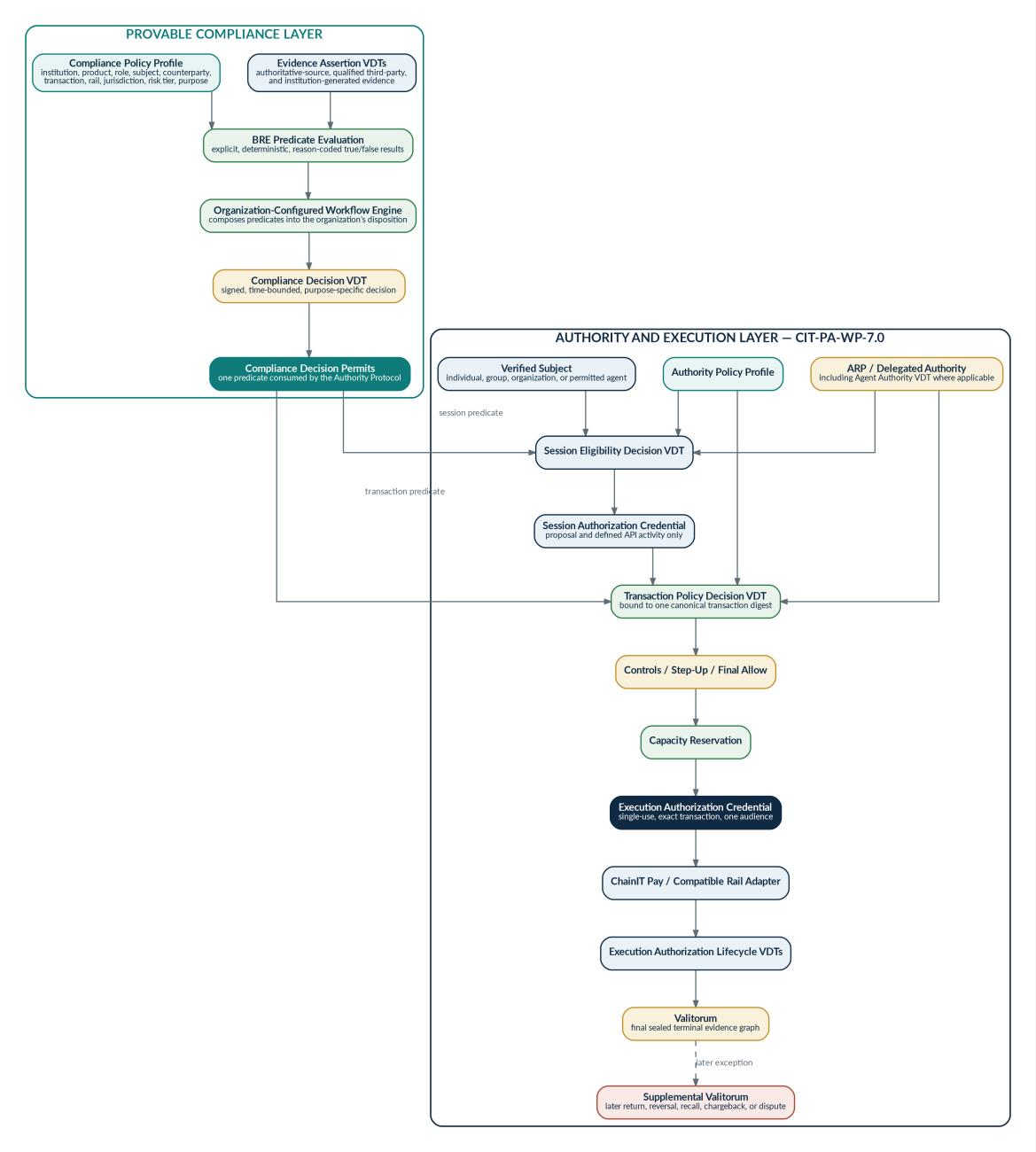}
  \caption{Provable Compliance within the final \href{https://chainit.com/insights/chainit-launches-provable-authority-as-u-s-generative-ai-enabled-fraud-losses-are-projected-to-reach-40-billion-by-2027/}{Authority Protocol}.}
  \label{fig:compliance-in-authority}
  \fignote{Compliance supplies a signed predicate; CIT-PA-WP-7.0 controls the session and transaction decision flows.}
\end{figure}
\clearpage %

\subsection{Session eligibility flow}\label{sec:session}

For a session that enables consequential activity, the CPP selects current compliance evidence and freshness requirements. The BRE returns reason-coded compliance predicates, and the Workflow Engine creates the Compliance Decision \href{https://chainit.com/product/data-tokenization/}{VDT}. The \href{https://chainit.com/insights/chainit-launches-provable-authority-as-u-s-generative-ai-enabled-fraud-losses-are-projected-to-reach-40-billion-by-2027/}{Authority Protocol} then evaluates Compliance Decision Permits together with subject authentication, the ARP or Delegated Authority VDT, the APP, status, scope, and any subject-specific inputs. Where the subject is an agent, the Session Eligibility Decision VDT additionally references the Agent Registration VDT, Agent Authority VDT, attested deployment, permitted tools, workload key, and ephemeral session-key hierarchy. \cite{ref23}

Only a final Allow session disposition may support SAC issuance. The SAC is short-lived and sender-constrained. It permits proposal activity, evidence retrieval, and defined API activity. Rail adapters must reject it as payment authority. An Allow with Controls, Step-Up, Review, Hold, Reject, Prohibit, or Stale/Unknown result follows the organization’s configured workflow and cannot silently fall through to session issuance.

\begin{figure}[H]
  \centering
  \includegraphics[width=453.57bp]{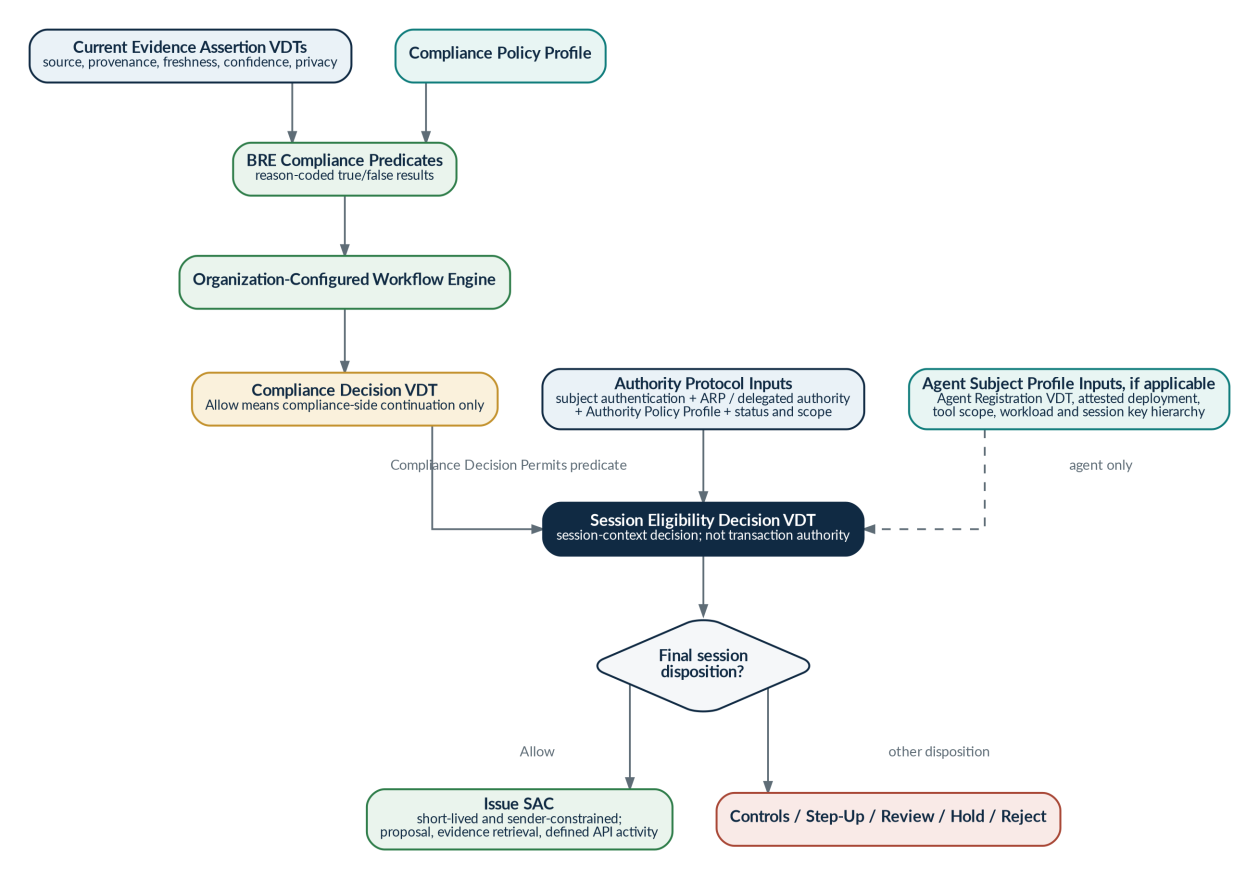}
  \caption{Compliance input to the Session Eligibility Decision \href{https://chainit.com/product/data-tokenization/}{VDT}.}
  \label{fig:session}
  \fignote{A compliance-side Allow permits continuation only; the \href{https://chainit.com/insights/chainit-launches-provable-authority-as-u-s-generative-ai-enabled-fraud-losses-are-projected-to-reach-40-billion-by-2027/}{Authority Protocol} still decides session eligibility.}
\end{figure}

\subsection{Transaction policy and commit-time flow}\label{sec:commit}

At the transaction reliance event, the Compliance Decision \href{https://chainit.com/product/data-tokenization/}{VDT} is evaluated with the ARP, Delegated Authority VDT or Agent Authority VDT, APP, verified subject, risk tier, canonical transaction digest, current capacity, required approvals, and rail-specific constraints. The resulting Transaction Policy Decision VDT is bound to one exact digest and one limited evaluation window. It is not reusable as standing status. \cite{ref23}

\begin{formulabox}{General execution predicate}
\formula{MayExecute = IdentityAssuranceSatisfied AND AuthorityValidUnderARP AND ComplianceDecisionPermits AND TransactionWithinScope AND RequiredApprovalsSatisfied AND BudgetAndRailControlsSatisfied}
\end{formulabox}

The payment-boundary predicate is the same test specialized to commit, where the exact instruction, reservation, status, final disposition, and one-time execution credential must also be verified:

\begin{formulabox}{Commit-time execution predicate}
\formula{MayCommit = SubjectAuthenticated AND AuthorityValidUnderARP AND EvidenceAcceptable AND ComplianceDecisionPermits AND WorkflowDispositionIsAllow AND StatusFresh AND CapacityReserved AND RequiredApprovalsSatisfied AND DigestMatches AND EACValidAndUnused}
\end{formulabox}

Every predicate is required. Allow with Controls or Step-Up must be completed and re-evaluated to final Allow before capacity reservation and EAC issuance. The reservation atomically holds capacity so concurrent requests cannot use the same limit twice. The EAC authorizes one exact transaction for one audience and is consumed atomically at the \href{https://chainit.com/product/digital-pay/}{ChainIT Pay} or compatible rail-adapter boundary. Failure produces a controlled disposition; there is no silent fallback to the SAC, a stale decision, or a generic bearer credential.

\bluehead{Commit sequence}\vspace{-1.4bp}
\setlength{\LTleft}{-57.3bp}
\tablepad{11.3bp}{4.8bp}
\begin{longtable}{|B{239.0bp}{7.5}{9}|M{353.0bp}{7.5}{9}|}
\hline
\headrow\thc{Gate} & \thc{Requirement}\\\hline
\endfirsthead
\hline
\headrow\thc{Gate} & \thc{Requirement}\\\hline
\endhead
1. Final decision & The Transaction Policy Decision \href{https://chainit.com/product/data-tokenization/}{VDT} must resolve to final Allow; a compliance-side Allow alone is insufficient.\\\hline
\altrow 2. Controls complete & Any Allow with Controls, Step-Up, Review, or Hold condition must be resolved and the transaction re-evaluated.\\\hline
3. Capacity reserved & The applicable limit or budget is atomically reserved so concurrent requests cannot commit the same capacity.\\\hline
\altrow 4. EAC verified & The one-time EAC must bind the exact digest, audience, reservation, status epoch, subject or session, and expiration, and remain unused.\\\hline
5. Commit and evidence & The rail boundary consumes the EAC, submits the exact instruction, and creates append-only lifecycle \href{https://chainit.com/product/data-tokenization/}{VDTs} leading to \href{https://chainit.com/}{Valitorum}.\\\hline
\end{longtable}

\begin{figure}[H]
  \centering
  \includegraphics[width=399.58bp]{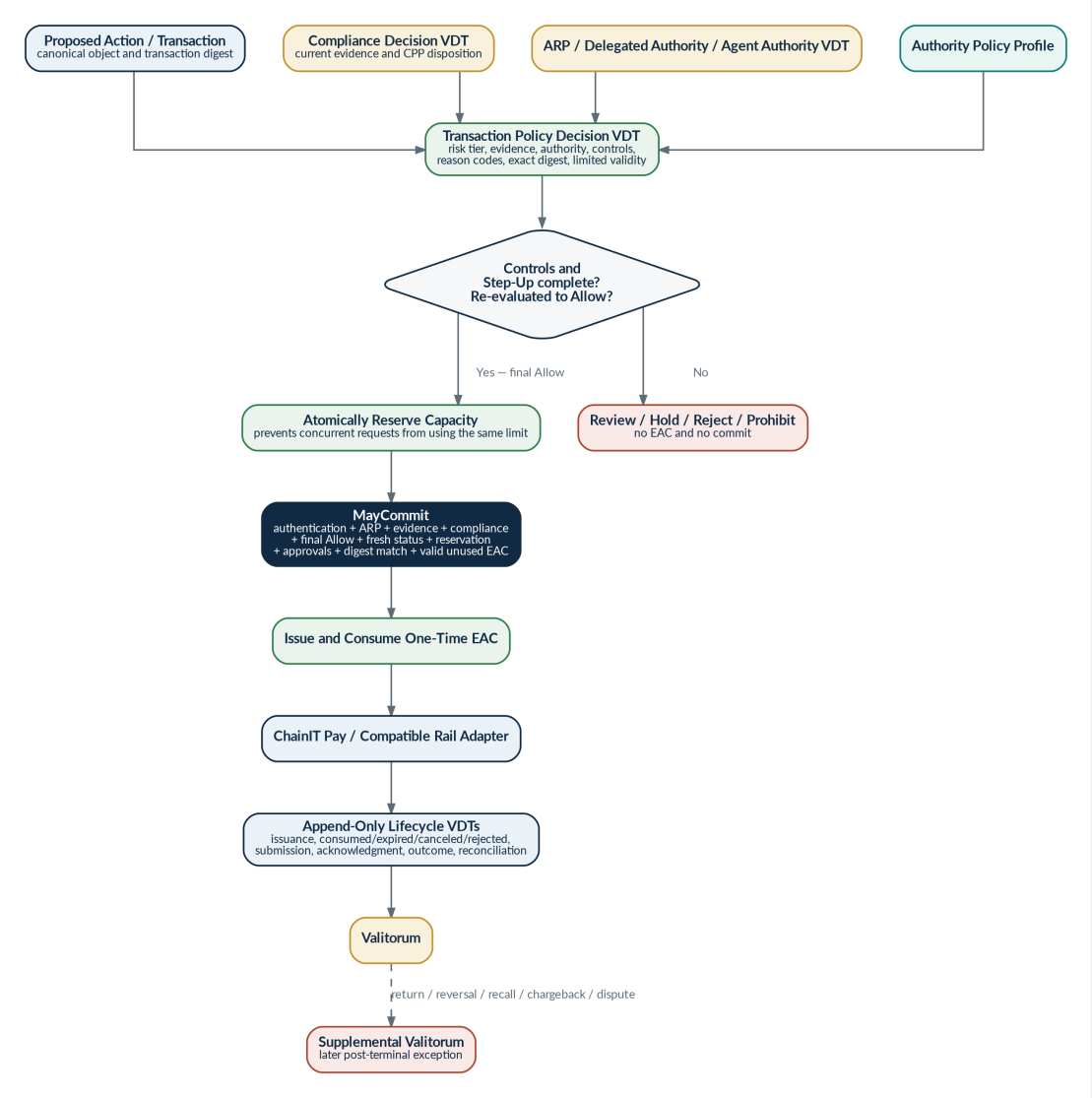}
  \caption{Compliance input to the transaction and commit-time flow.}
  \label{fig:commit}
  \fignote{The Compliance Decision \href{https://chainit.com/product/data-tokenization/}{VDT} contributes one predicate; only final Allow plus reservation and a valid unused EAC can commit value.}
\end{figure}

\subsection{Agent Subject Profile boundary}\label{sec:agent}

Only the Agent Registration \href{https://chainit.com/product/data-tokenization/}{VDT}, the rotatable workload key, the ephemeral session key, and the sovereignty boundary are Agent Subject Profile artifacts. The ARP, Delegated Authority VDT, Agent Authority VDT, Evidence Assertion VDT, Compliance Decision VDT, Session Eligibility Decision VDT, SAC, Transaction Policy Decision VDT, capacity reservation, EAC, Execution Authorization Lifecycle VDTs, \href{https://chainit.com/}{Valitorum}, and Supplemental Valitorum are protocol-general. Their inputs vary by subject class; their authority and execution meaning does not. \cite{ref23}

Agent registration, workload attestation, and session-key containment are inputs supplied by the Agent Subject Profile. \href{https://chainit.com/white-papers/}{Provable Compliance} may evaluate their current evidence state—for example, whether the registered deployment or key status remains acceptable—but it does not redefine them, and they do not themselves confer authority. An AI model may reason, gather evidence, and propose an action. It does not receive unrestricted payment credentials, custody keys, or final execution authority.

\subsection{Pactvera, lifecycle VDTs, Valitorum, and later exceptions}\label{sec:pactvera}

Where a compliance decision governs a \href{https://www.pactvera.com/}{Pactvera} transaction, the CPP version, Evidence Assertion \href{https://chainit.com/product/data-tokenization/}{VDT} references, predicate results, Compliance Decision VDT, review or hold case references, disclosure classification, and material state transitions become part of the same append-only evidence lifecycle used by the \href{https://chainit.com/insights/chainit-launches-provable-authority-as-u-s-generative-ai-enabled-fraud-losses-are-projected-to-reach-40-billion-by-2027/}{Authority Protocol}. Pactvera and the \href{https://chainit.com/}{Valitorum} Service provide the transaction logic and terminal evidence package. \cite{ref18,ref19,ref23}

The operational EAC is short-lived and is not stored as a reusable ledger object. Execution Authorization Lifecycle \href{https://chainit.com/product/data-tokenization/}{VDTs} record issuance and terminal treatment. Submission, acknowledgment, settlement, rejection, cancellation, return, reversal, recall, chargeback, dispute, and reconciliation are separate linked events; an earlier immutable event is never rewritten. The \href{https://chainit.com/}{Valitorum} seals the completed or otherwise terminal evidence graph under a defined completeness rule. A later post-terminal exception may be sealed in a Supplemental Valitorum referencing the original.

\addtocontents{toc}{\protect\setcounter{tocdepth}{1}}%
\section{Runtime Re-Verification at the Point of Reliance}\label{sec:runtime}

The defining commitment of \href{https://chainit.com/white-papers/}{Provable Compliance} is point-of-reliance evaluation. ChainIT does not assume that an onboarding result remains valid forever. It also does not impose the technically and economically unrealistic requirement that every source be queried for every event. The system evaluates whether each required item remains fresh enough for the present context and refreshes it when source volatility, transaction risk, policy, or a material event requires a new observation.

\subsection{Reliance triggers}\label{sec:triggers}

\begin{enumerate}
  \item Session issuance or renewal for an individual, group, organizational representative, or agent where the session enables consequential action.
  \item A transaction proposal, payment initiation, smart-contract execution, or other high-impact action.
  \item A human step-up approval, particularly for a new payee, changed destination, higher value, irreversible rail, elevated risk, or required quorum.
  \item A scheduled review established by source volatility, product risk, jurisdiction, relationship risk, or institutional policy.
  \item A source-change notification, sanctions-list update, license change, credential-expiry event, ownership change, AML alert, policy update, ARP or delegated-authority change, key or registration change, or another material-risk event.
\end{enumerate}

\subsection{Freshness, not universal live lookup}\label{sec:freshness}

For each required evidence item, the CPP establishes a maximum acceptable age and any conditions requiring live verification. The BRE evaluates the evidence against the current reliance event:

\begin{formulabox}{Freshness rule}
\formula{FreshEnough = (Current Time − Evidence Observed Time) ≤ Maximum Age permitted for the source, evidence type, risk tier, product, jurisdiction, and reliance event}
\end{formulabox}

A live source interaction may be required for high-risk or legally sensitive activity. A recently signed Evidence Assertion \href{https://chainit.com/product/data-tokenization/}{VDT} may be sufficient for a lower-risk action when policy permits. Event-driven invalidation shortens the acceptable interval. If a source is unavailable, the system applies the approved degraded-mode rule—such as stronger verification, Review, Hold, Stale/Unknown, or fail-closed—rather than silently assuming that a prior result remains current. This source-specific model also reflects the risk-based nature of current U.S. CDD guidance. \cite{ref1}

\subsection{Status epoch and verification rigor}\label{sec:epoch}

The \href{https://chainit.com/insights/chainit-launches-provable-authority-as-u-s-generative-ai-enabled-fraud-losses-are-projected-to-reach-40-billion-by-2027/}{Authority Protocol} uses status epochs or status-freshness boundaries to determine whether cached non-revocation and evidence state remain acceptable for a risk tier. Tier 1–Low may permit a signed, recent status snapshot where the APP and CPP allow it. Tier 2–Standard may require a shorter freshness window or targeted live checks. Tier 3–High may require live status introspection or live source verification. The organization’s Workflow Engine assigns the risk tier and records the policy version, factors, time, reason codes, and any manual override; the BRE only evaluates the configured predicates. \cite{ref23}

\subsection{Scoped propagation and revocation}\label{sec:scope}

A change in evidence propagates to the scope it actually affects. A confirmed sanctions prohibition may require immediate blocking of the affected person, entity, property, wallet, or transaction. A potential name match requires Review or Hold. A PEP designation may trigger enhanced diligence. An expired professional license blocks the licensed activity. A source outage may hold high-risk activity while allowing lower-risk, previously verified activity under an approved policy. The affected-scope calculation identifies the subject, organization, product, role, agent, session, transaction class, counterparty, destination, jurisdiction, and authority relationships that depend on the changed evidence. Broader suspension is used only where law, source state, or configured policy supports it.

\begin{figure}[H]
  \centering
  \includegraphics[width=320.38bp]{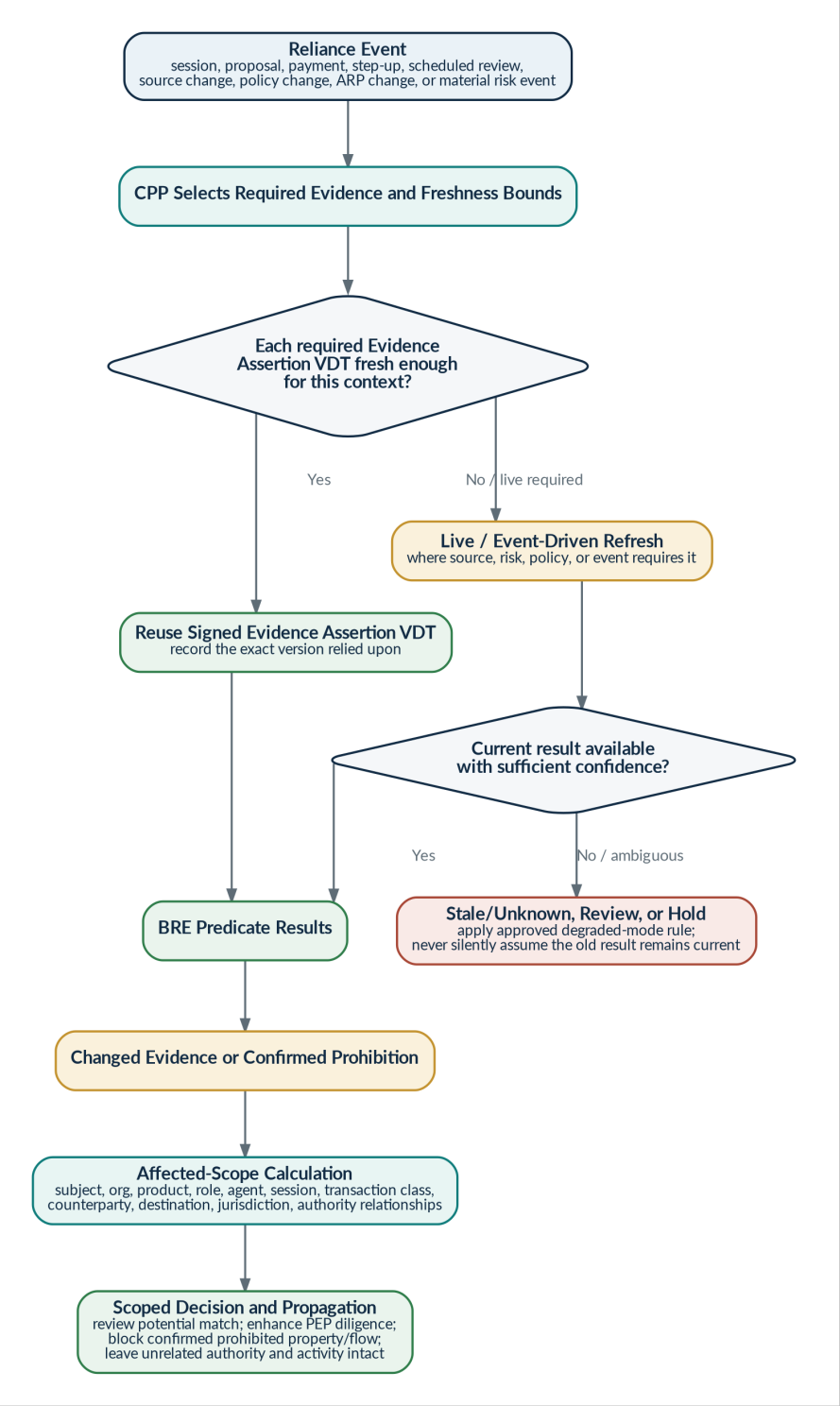}
  \caption{Runtime freshness and affected-scope decisioning.}
  \label{fig:freshness}
  \fignote{Current evidence is reused only within the applicable freshness bound; changed evidence propagates to the scope it actually affects.}
\end{figure}

\section{Evidence Domains}\label{sec:domains}

Evidence Assertion \href{https://chainit.com/product/data-tokenization/}{VDTs} are not limited to a fixed catalog. Any sourced or institution-generated fact relevant to a compliance decision may be represented under the same architecture. The domains below recur often enough to address in detail, but they are not equivalent Boolean gates and they do not produce one universal compliance badge.

\setlength{\LTleft}{-28.8bp}
\tablepad{12.3bp}{7.0bp}
\begin{longtable}{|B{155.9bp}{7.5}{10.15}|M{212.2bp}{7.5}{10.15}|M{155.9bp}{7.5}{10.15}|}
\hline
\headrow\thc[7.5]{Evidence domain} & \thc[7.5]{What it represents} & \thc[7.5]{Typical decision effect}\\\hline
\endfirsthead
\hline
\headrow\thc[7.5]{Evidence domain} & \thc[7.5]{What it represents} & \thc[7.5]{Typical decision effect}\\\hline
\endhead
AML risk and monitoring & Institution-generated customer, relationship, transaction, source-of-funds, alert, case, and monitoring state. There is no universal external “AML-clear” status. & Allow, enhanced monitoring, Step-Up, Review, Hold, Reject, reporting, or another institution-specific disposition.\\\hline
\altrow OFAC and broader sanctions & Official list data plus ownership/control analysis, program restrictions, geography, wallet/address indicators, match confidence, and block-versus-reject rules. & Potential match → Review or Hold. Confirmed applicable prohibition → Prohibit or block, with required reporting and recordkeeping.\\\hline
PEP and relationship risk & Whether a subject is or is associated with a PEP, together with office, geography, relationship, source-of-wealth/funds, and mitigating evidence. & Enhanced diligence, management approval, monitoring, controls, or risk-based allowance—not automatic universal rejection.\\\hline
\altrow Criminal or background evidence & Purpose-limited reports relevant to a role, license, employment, housing, credit, regulated activity, or another lawful use. & Applied only where lawful and relevant, with required consent, notice, accuracy, dispute, individualized-assessment, and adverse-action processes.\\\hline
Government identity and assurance & Identity-source match, issuing jurisdiction, credential state where available, \href{https://chainit.com/product/identity-verification/}{ChainIT ID} lifecycle, biometric reauthentication, device context, and assurance. & May require identity refresh or stronger authentication. Expiration of one credential does not erase identity or corporate authority.\\\hline
\altrow Jurisdiction, product, licensing, KYC/KYB, and payment transparency & Product- and location-specific obligations, licenses, entity status, customer or beneficial-owner evidence, Travel Rule data, and sector controls. & Selects policy and evidence; limits only the implicated activity, product, role, or geography unless a broader rule applies.\\\hline
\end{longtable}

\subsection{AML risk and monitoring}\label{sec:aml}

AML monitoring is a process performed by the responsible institution, not a government status feed. It evaluates patterns, relationships, expected activity, counterparties, geography, source and destination of funds, products, velocity, structuring indicators, internal alerts, and investigative findings. An Evidence Assertion \href{https://chainit.com/product/data-tokenization/}{VDT} may represent the institution’s current risk and case state, the model or rule version, and the permitted disclosure level. U.S. AML/CFT requirements remain risk-based, and federal agencies continued modernizing that framework through rulemaking in 2026. \cite{ref1,ref2}

\subsection{OFAC and broader sanctions}\label{sec:ofac}

Sanctions screening is not limited to checking whether a named person appears on the SDN List. OFAC obligations may extend to entities owned 50 percent or more, directly or indirectly and in the aggregate, by blocked persons; program-specific prohibitions; prohibited trade or investment; digital-currency addresses and other identifiers; and transactions that evade, avoid, cause, or attempt a violation. OFAC states that digital-currency and fiat transactions are subject to the same sanctions obligations. \cite{ref5,ref6} The evidence model therefore preserves list version, retrieval time, ownership and control analysis, identifier and wallet results, match confidence, disposition basis, and required reporting or recordkeeping.

\subsection{PEP status and enhanced due diligence}\label{sec:pep}

PEP status is a risk indicator, not a finding of illegality. U.S. banking agencies state that PEP relationships present varying levels of risk and do not automatically require one unique procedure merely because a customer is considered a PEP. FATF similarly describes PEP controls as preventive rather than criminal. \cite{ref3,ref4} The policy response may include source-of-funds or source-of-wealth evidence, senior approval, enhanced monitoring, or other proportionate controls.

\subsection{Criminal and background evidence}\label{sec:background}

Criminal-history and broader background information is enabled only where the CPP identifies a lawful, relevant purpose and the required consent, notice, permissible-purpose, accuracy, dispute, nondiscrimination, and adverse-action controls. Where an employer obtains a report from a consumer-reporting company, the FCRA and related employment requirements may apply. \cite{ref8} Raw records are not placed in a broadly reusable Compliance Decision \href{https://chainit.com/product/data-tokenization/}{VDT}.

\subsection{Government identity and assurance freshness}\label{sec:identity}

The identity materials support government-source verification, biometric authentication, selective disclosure, lifecycle updates, and later verification without accessing the government system every time. \cite{ref17} AAMVA’s Driver’s License Data Verification service allows approved users to submit driver’s-license or identification data and receive match flags against participating issuing-jurisdiction records; it does not establish universal coverage or prove every legal attribute a relying party may require. \cite{ref7} The decision therefore distinguishes identity, credential state, driving privilege, assurance freshness, and the transaction policy requiring a current credential.

\subsection{Jurisdiction-, product-, and industry-specific evidence}\label{sec:jurisdiction}

Licensing, registration, KYC/KYB, beneficial ownership, age, professional qualification, Travel Rule, gaming, healthcare, employment, environmental, safety, insurance, tax, entity-standing, and other obligations vary materially by activity and jurisdiction. The CPP identifies the applicable regulated entity, product, role, rail, transaction type, and location. Adding a domain does not require a new architecture; it requires an appropriate source class, Evidence Assertion \href{https://chainit.com/product/data-tokenization/}{VDT} schema, freshness rule, predicate set, workflow configuration, and disclosure policy.

\section{Regulatory Landscape and Current Status}\label{sec:regulatory}

\goldpanel{Regulatory currency}{This section reflects the official sources listed in this edition and is stated as current through August 31, 2026. Rules, effective dates, agency positions, and national implementation continue to change. The section supports architecture and general discussion; it is not product- or jurisdiction-specific legal advice.}

\subsection{U.S. AML/BSA and customer due diligence}\label{sec:cdd}

The U.S. CDD framework requires covered financial institutions to understand the nature and purpose of customer relationships, conduct ongoing monitoring to identify and report suspicious transactions, and maintain and update customer information on a risk basis. FinCEN’s May 2026 consolidated FAQs emphasize that there is no categorical requirement to update customer information on one continuous or periodic schedule; risk-related information and ordinary monitoring determine when reassessment is required. \cite{ref1} In February 2026, FinCEN also granted exceptive relief from re-identifying and re-verifying beneficial owners each time a legal-entity customer opens another account, while preserving initial verification, risk-triggered updating, and other ongoing obligations. \cite{ref1}

\subsection{OFAC and sanctions}\label{sec:ofac2}

OFAC expects sanctions controls tailored to the business and its risk. The architecture can preserve screening of relevant individuals, entities, ownership and control, counterparties, wallets or identifiers, geography, and program restrictions; possible-match resolution; block-versus-reject treatment; and the evidence needed for reporting and recordkeeping. Digital-asset transactions do not receive a lower sanctions standard because they use a different rail. \cite{ref5,ref6}

\subsection{Stablecoin-specific requirements under the GENIUS Act}\label{sec:genius}

The GENIUS Act was enacted July 18, 2025 as Public Law 119-27. \cite{ref10} It establishes a federal and state framework for permitted payment-stablecoin issuers, foreign issuers, custody, reserves, redemption, supervision, and related activities. The Act’s effective date is the earlier of eighteen months after enactment or 120 days after the primary federal regulators issue final implementing regulations. The OCC’s February 2026 proposal described January 18, 2027 as the expected date absent an earlier trigger. \cite{ref14}

As of August 31, 2026, implementation remained subject to active rulemaking. FinCEN and the federal banking agencies had proposed customer-identification requirements for permitted payment-stablecoin issuers; FinCEN and OFAC had jointly proposed broader AML/CFT and sanctions-program requirements; and Treasury issued a separate Section 3 proposed rule on August 17, 2026 concerning issuance, offering, and sale of payment stablecoins in the United States. \cite{ref11,ref12,ref13} \href{https://chainit.com/white-papers/}{Provable Compliance} can help an issuer, bank, wallet, or service provider evidence screening and policy decisions. It does not itself determine whether ChainIT or another party is an issuer, custodian, money transmitter, digital-asset service provider, or other regulated actor. That perimeter depends on actual issuance, redemption, custody, control, transfer power, conversion, and service functions.

\subsection{FATF Travel Rule and cross-border payment transparency}\label{sec:travel}

Verified identity records can supply originator and beneficiary data, but storing the data is not the same as satisfying the Travel Rule. Responsible entities must obtain, validate where required, transmit or make available, retain, protect, and screen applicable information under national law and interoperable messaging arrangements. FATF Recommendation 15 and its interpretive materials continue to govern virtual assets and VASPs; FATF’s July 2026 update reported increasing implementation together with continuing operational and supervisory gaps. \cite{ref15} FATF strengthened Recommendation 16 in June 2025 and conducted a public consultation on implementation guidance from June 24 through August 21, 2026, with countries expected to be ready for revised requirements by the end of 2030. \cite{ref16}

\subsection{Identity, background, licensing, and sector rules}\label{sec:sector}

Identity, background, licensing, and registration requirements vary by purpose and jurisdiction. AAMVA DLDV is a valuable U.S. source service but depends on approved access and participating jurisdictions. \cite{ref7} Background reports may implicate FCRA, employment, housing, credit, licensing, consumer-reporting, and local restrictions. \cite{ref8} Professional licenses and entity registrations require their own official or qualified sources and status semantics. The CPP selects those requirements; no one Evidence Assertion \href{https://chainit.com/product/data-tokenization/}{VDT} or Compliance Decision VDT is a universal global compliance badge.

\section{Risk Tier, Decision States, and Enforcement}\label{sec:states}

Risk tier and disposition are independent. The organization’s Workflow Engine assigns Tier 1–Low, Tier 2–Standard, or Tier 3–High based on configured factors and records the policy version, inputs, assignment time, reason codes, and any manual override. The risk tier selects minimum verification rigor. The BRE returns true/false predicates. The organization’s Workflow Engine then combines those predicates into the disposition. A high-risk transaction may resolve to Allow after required controls are satisfied; a low-risk transaction may be rejected because a definitive policy predicate failed. \cite{ref23}

\setlength{\LTleft}{-37.8bp}
\tablepad{12.3bp}{7.0bp}
\begin{longtable}{|B{155.9bp}{7.5}{10.15}|M{230.2bp}{7.5}{10.15}|M{155.9bp}{7.5}{10.15}|}
\hline
\headrow\thc[7.5]{Disposition} & \thc[7.5]{Compliance-side meaning} & \thc[7.5]{Typical next action}\\\hline
\endfirsthead
\hline
\headrow\thc[7.5]{Disposition} & \thc[7.5]{Compliance-side meaning} & \thc[7.5]{Typical next action}\\\hline
\endhead
Allow & All compliance-side predicates are satisfied for the stated scope and time. & Continue to the Session Eligibility or Transaction Policy Decision; do not issue an EAC solely from this result.\\\hline
\altrow Allow with Controls & Continuation is permitted subject to stated restrictions, monitoring, limits, reporting, or safeguards. & Apply controls, record completion, and re-evaluate to final Allow where the \href{https://chainit.com/insights/chainit-launches-provable-authority-as-u-s-generative-ai-enabled-fraud-losses-are-projected-to-reach-40-billion-by-2027/}{Authority Protocol} requires it.\\\hline
Step-Up & A stronger identity event, human approval, second approver, additional source, or enhanced diligence is required. & Pause and initiate the configured step-up workflow.\\\hline
\altrow Review & A potential match, elevated risk, conflicting evidence, or contextual judgment requires an authorized reviewer. & Open or reference a case; preserve confidentiality and reviewer authority.\\\hline
Hold & Required evidence is unavailable, incomplete, stale, or unresolved. & Do not execute; obtain evidence, restore source access, or apply approved degraded mode.\\\hline
\altrow Reject & The action violates institution policy, risk appetite, or authority conditions but is not necessarily legally prohibited. & Decline with the audience-appropriate response.\\\hline
Prohibit & A mandatory legal or regulatory prohibition applies. & Block or reject as required, preserve records, and perform required reporting.\\\hline
\altrow Stale/Unknown & The system cannot support a current decision within required freshness and confidence bounds. & Fail controlled; Hold, Review, or deny according to the applicable profile.\\\hline
\end{longtable}

The Compliance Decision \href{https://chainit.com/product/data-tokenization/}{VDT} records the compliance-side disposition. That disposition does not become the final \href{https://chainit.com/insights/chainit-launches-provable-authority-as-u-s-generative-ai-enabled-fraud-losses-are-projected-to-reach-40-billion-by-2027/}{Authority Protocol} decision by implication. For sessions, it is consumed by the Session Eligibility Decision VDT. For transactions, it is consumed by the Transaction Policy Decision VDT. Only final Allow after all required controls, approvals, status checks, scope checks, digest checks, capacity reservation, and credential checks can satisfy MayCommit. \cite{ref23}

\section{Design Principles, Evidence and Decision Lifetimes, and Residual Risk}\label{sec:principles}

\subsection{Zero-Trust Design Principles}\label{sec:zero-trust}

\begin{itemize}
  \item Sourced and attributable. Every evidence item identifies who produced it, which source class applies, what was observed, when it was observed, and what it does and does not prove.
  \item Contextual rather than universal. The decision is tied to an institution, product, role, subject, counterparty, transaction, rail, jurisdiction, risk tier, and purpose.
  \item Fresh enough for the present risk. Evidence is not trusted indefinitely, but sources are not queried unnecessarily. Freshness policy is explicit and measurable.
  \item Authority remains separate. The ARP records organizational power; Delegated Authority and Agent Authority \href{https://chainit.com/product/data-tokenization/}{VDTs} expose bounded grants; the Compliance Decision VDT gates present exercise without rewriting any grant.
  \item Compliance is one predicate, not payment authority. A favorable result permits continuation into the final session or transaction decision and cannot issue or replace an EAC.
  \item Risk tier is not disposition. Verification rigor and decision outcome are separately determined and recorded.
  \item Uncertainty is a state. Possible matches, source outages, stale evidence, and incomplete data produce Review, Hold, or Stale/Unknown rather than silent allowance or indiscriminate permanent blocking.
  \item Least disclosure. Subjects and counterparties receive only the minimum outcome; authorized institutional personnel and lawful oversight recipients receive deeper evidence according to role and law.
  \item Append-only evidence. Refreshes, reviews, approvals, decisions, EAC lifecycle states, outcomes, and later exceptions create new \href{https://chainit.com/product/data-tokenization/}{VDTs} or Supplemental Valitorums. A sealed record is never edited.
  \item Human accountability. Overrides, manual findings, approvals, reviews, and release decisions identify the responsible verified person, governing authority, reason, time, and scope.
  \item Resilience and recovery. Source outages, false positives, key compromise, data corrections, appeals, policy changes, rail ambiguity, and later returns have defined controlled paths.
  \item Agent containment. Agent registration and keys identify and contain the workload, but authority still comes from the ARP and the agent cannot cross the sovereignty boundary to unrestricted payment credentials or final execution authority.
\end{itemize}

\subsection{Evidence and Decision Lifetimes}\label{sec:lifetimes}

\href{https://chainit.com/white-papers/}{Provable Compliance} artifacts inherit the platform foundation and \href{https://chainit.com/insights/chainit-launches-provable-authority-as-u-s-generative-ai-enabled-fraud-losses-are-projected-to-reach-40-billion-by-2027/}{Authority Protocol} control boundaries described in CIT-PA-WP-7.0. The compliance-specific artifacts below are versioned or reliance-event-specific; none is a permanent status, and a later reviewer can determine exactly which evidence, policy, and decision were current at the time of reliance.

\tablehead{Evidence and Policy Artifacts}\vspace{-1.4bp}
\setlength{\LTleft}{13.7bp}
\tablepad{11.3bp}{4.8bp}
\begin{longtable}{|B{99.2bp}{7.5}{9}|M{64.1bp}{7.5}{9}|M{154.2bp}{7.5}{9}|M{110.6bp}{7.5}{9}|}
\hline
\headrow\thc{Artifact} & \thc{Layer} & \thc{Typical lifetime or state} & \thc{Control boundary}\\\hline
\endfirsthead
\hline
\headrow\thc{Artifact} & \thc{Layer} & \thc{Typical lifetime or state} & \thc{Control boundary}\\\hline
\endhead
Evidence Assertion \href{https://chainit.com/product/data-tokenization/}{VDT} & Compliance layer & Source-specific effective and expiration periods; refreshed or re-derived at the reliance event rather than trusted indefinitely. & Evidence service; selective disclosure.\\\hline
\altrow Compliance Policy Profile (CPP) & Compliance layer & Versioned; effective until amended, superseded, or retired. & Institution’s compliance policy configuration.\\\hline
Compliance Decision \href{https://chainit.com/product/data-tokenization/}{VDT} & Compliance layer & Signed, time-bounded, reliance-event-specific; superseded by a new decision rather than altered. & Organization’s configuration of the ChainIT Workflow Engine.\\\hline
\altrow Compliance Decision Permits (predicate) & Compliance layer & Evaluated fresh at each reliance event; not cached or reused as a standing status. & \href{https://chainit.com/insights/chainit-launches-provable-authority-as-u-s-generative-ai-enabled-fraud-losses-are-projected-to-reach-40-billion-by-2027/}{Authority Protocol} / BRE.\\\hline
\end{longtable}

\tablehead{Decision, Transaction, and Evidence-Graph Artifacts}\vspace{-1.4bp}
\tablepad{11.3bp}{4.8bp}
\begin{longtable}{|B{99.2bp}{7.5}{9}|M{64.1bp}{7.5}{9}|M{154.2bp}{7.5}{9}|M{110.6bp}{7.5}{9}|}
\hline
\headrow\thc{Artifact} & \thc{Layer} & \thc{Typical lifetime or state} & \thc{Control boundary}\\\hline
\endfirsthead
\hline
\headrow\thc{Artifact} & \thc{Layer} & \thc{Typical lifetime or state} & \thc{Control boundary}\\\hline
\endhead
Session Eligibility Decision \href{https://chainit.com/product/data-tokenization/}{VDT} (compliance input) & \href{https://chainit.com/insights/chainit-launches-provable-authority-as-u-s-generative-ai-enabled-fraud-losses-are-projected-to-reach-40-billion-by-2027/}{Authority Protocol} & One session context; expires and is re-derived, consuming the current Compliance Decision Permits result. & Authority gateway / BRE.\\\hline
\altrow Transaction Policy Decision \href{https://chainit.com/product/data-tokenization/}{VDT} (compliance input) & \href{https://chainit.com/insights/chainit-launches-provable-authority-as-u-s-generative-ai-enabled-fraud-losses-are-projected-to-reach-40-billion-by-2027/}{Authority Protocol} & One exact transaction and limited evaluation window; not reusable as standing status. & BRE / transaction control plane.\\\hline
Execution Authorization Lifecycle \href{https://chainit.com/product/data-tokenization/}{VDT} & \href{https://chainit.com/insights/chainit-launches-provable-authority-as-u-s-generative-ai-enabled-fraud-losses-are-projected-to-reach-40-billion-by-2027/}{Authority Protocol} & Permanent immutable evidence events for issuance and terminal handling. & ChainIT evidence ledger.\\\hline
\altrow \href{https://chainit.com/}{Valitorum} & Platform foundation & Permanent immutable seal over the configured terminal evidence graph, including compliance references. & Valitorum verification package.\\\hline
Supplemental \href{https://chainit.com/}{Valitorum} & \href{https://chainit.com/insights/chainit-launches-provable-authority-as-u-s-generative-ai-enabled-fraud-losses-are-projected-to-reach-40-billion-by-2027/}{Authority Protocol} & Permanent seal over a later exception graph (return, reversal, recall, chargeback, dispute, or later compliance finding) referencing the original Valitorum. & Post-terminal exception verification package.\\\hline
\end{longtable}

\global\tablelastfalse\vspace{-8.6bp}%
\subsection{Residual Risk and Non-Goals}\label{sec:residual}

\href{https://chainit.com/white-papers/}{Provable Compliance} is designed to make evidence, policy selection, predicate evaluation, disposition, and reliance-event context attributable and independently verifiable. It does not eliminate:

\begin{itemize}
  \item Incorrect, delayed, or incomplete data from a sanctions list, PEP database, credit bureau, government registry, or other third-party evidence source.
  \item A false negative or false positive produced by a screening, matching, or scoring algorithm operating outside ChainIT’s control.
  \item An institution configuring its own Compliance Policy Profile or disposition set in a way that does not satisfy its actual regulatory obligations.
  \item Coercion, collusion, or malicious conduct by people who satisfy the formal evidence and policy rules.
  \item Compromise of ChainIT issuer infrastructure, an evidence provider, or a source system outside ChainIT’s control.
  \item Residual exposure accepted by a verifier that relies on a signed evidence snapshot up to the approved maximum staleness.
  \item A regulatory change, guidance update, or enforcement action that has not yet been reflected in the applicable CPP.
\end{itemize}

\redpanel{Claims discipline}{Provable Compliance is designed to reduce stale, mismatched, undocumented, and non-attributable compliance decisions. It is not a guarantee of regulatory compliance, a substitute for an institution’s AML/CFT or sanctions program, or a certification that a subject is free of risk, and it is not described as such. The residual risks above are stated so that a deploying institution can control for them.}

\section{Illustrative Scenarios}\label{sec:scenarios}

\subsection{Human-initiated wire under the same ARP}\label{sec:s-wire}

Jane, Acme Corp’s finance director, authenticates through \href{https://chainit.com/product/identity-verification/}{ChainIT ID} and proposes a \$60,000 wire under an ARP authorizing treasury activity up to \$100,000 per transaction, subject to a second authorized approver above \$25,000, within specified counterparties, destinations, and approval rules. The CPP selects sanctions, customer, counterparty, jurisdiction, and other required evidence. The BRE returns deterministic compliance predicates, and the Workflow Engine creates a Compliance Decision \href{https://chainit.com/product/data-tokenization/}{VDT}. That result becomes the Compliance Decision Permits input to the Transaction Policy Decision VDT. The transaction still must satisfy the ARP, APP, canonical digest, required approvals, capacity, status, and rail conditions before final Allow, reservation, and EAC issuance. A compliance-side Allow alone cannot release the wire.

\subsection{Agent-initiated payment within delegated scope}\label{sec:s-agent}

Acme Corp’s procurement agent is registered under the Agent Subject Profile. Jane, acting under the ARP, grants a bounded Agent Authority \href{https://chainit.com/product/data-tokenization/}{VDT} permitting approved cloud-service payments up to \$5,000 per transaction and \$25,000 per month. The agent authenticates its registered deployment and receives a SAC after the Session Eligibility Decision VDT resolves to final Allow. The SAC permits invoice review and proposal activity but is rejected as payment authority. For a \$3,800 payment, the compliance layer returns an Allow, the Transaction Policy Decision VDT confirms the exact digest and scope, capacity is reserved, and a one-time EAC is consumed by \href{https://chainit.com/product/digital-pay/}{ChainIT Pay}. The lifecycle events and final outcome are sealed in the \href{https://chainit.com/}{Valitorum}.

\subsection{Quorum and separation-of-duties transaction}\label{sec:s-quorum}

An organization-controlled ACH file requires two approvers under the ARP. Each approver is a subject of authority, and each may have a separate Compliance Decision \href{https://chainit.com/product/data-tokenization/}{VDT} based on identity assurance, role-relevant evidence, and current policy. The group does not become a new identity merely because a quorum is required. The Transaction Policy Decision VDT records the quorum rule and exact digest. Capacity is not committed and no EAC is issued until both approvals refer to the same digest and all compliance and authority predicates are final Allow.

\subsection{Confirmed sanctions prohibition affecting a vendor}\label{sec:s-sanctions}

A vendor becomes subject to an applicable sanctions prohibition. At the next reliance event—or earlier through a source-change feed—the sanctions Evidence Assertion \href{https://chainit.com/product/data-tokenization/}{VDT} is refreshed. The BRE evaluates the relevant predicates, and Acme’s Workflow Engine issues Prohibit for the affected counterparty and flows. Jane’s procurement authority, the agent’s grant, and unrelated vendors remain intact unless the applicable rule requires broader blocking. The system records the source version, ownership or identifier analysis, decision, affected scope, and reporting treatment without exposing investigation detail to the vendor.

\subsection{Newly identified PEP requiring enhanced diligence}\label{sec:s-pep}

Jane is appointed to a public position and is treated as a PEP under the applicable CPP. Her corporate role and ARP do not disappear. The changed Evidence Assertion \href{https://chainit.com/product/data-tokenization/}{VDT} causes Review or Allow with Controls. Acme may require source-of-funds or source-of-wealth information, senior approval, enhanced monitoring, or a lower autonomous threshold on grants she made. If the controls are completed, a new Compliance Decision VDT may permit continuation and the \href{https://chainit.com/insights/chainit-launches-provable-authority-as-u-s-generative-ai-enabled-fraud-losses-are-projected-to-reach-40-billion-by-2027/}{Authority Protocol} re-evaluates the session or transaction. The record preserves the difference between a risk factor and a prohibition.

\subsection{Identity credential no longer meeting the assurance profile}\label{sec:s-credential}

Jane’s driver’s license expires. ChainIT continues to know which verified person the \href{https://chainit.com/product/identity-verification/}{ChainIT ID} represents, but the government-credential evidence no longer satisfies Acme’s required assurance for a high-value biometric step-up. The compliance result is Step-Up, Hold, or Stale/Unknown according to the CPP; it is not a finding that Jane ceased to exist or lost every corporate role. Jane refreshes ChainIT ID using an accepted current credential or approved alternative process. A new evidence \href{https://chainit.com/product/data-tokenization/}{VDT} and Compliance Decision VDT are created, and the same ARP and transaction digest are re-evaluated.

\subsection{Later return or dispute after terminal settlement}\label{sec:s-return}

A payment settles and the configured completeness rule is satisfied, so the \href{https://chainit.com/}{Valitorum} is sealed. A later ACH return, stablecoin reversal mechanism, recall, chargeback, or dispute does not alter the original Valitorum. The rail outcome and any new compliance or review decision are recorded as new linked lifecycle \href{https://chainit.com/product/data-tokenization/}{VDTs} and may be sealed in a Supplemental Valitorum referencing the original. This preserves both the original terminal record and the later exception history.

\section{Privacy, Disclosure, and Evidentiary Controls}\label{sec:privacy}

A system capable of explaining why a compliance decision changed must also prevent unnecessary or prohibited disclosure. The Compliance Decision \href{https://chainit.com/product/data-tokenization/}{VDT} separates verifiable outcome data from restricted evidence, case notes, raw source records, and legally protected information. The disclosure tier is selected by policy and audience, and every access or export should itself be attributable and auditable.

Independent verification establishes the issuer, integrity, policy version, evidence references, timestamps, status, and lifecycle of the decision artifact. It does not necessarily permit a third party to re-query a restricted source or independently establish every underlying source fact. For public records or source-signed evidence, a relying party may be able to reproduce or independently validate the underlying fact directly. For restricted government data, commercial screening services, SAR-related information, or institution-generated monitoring, verification is limited to the decision artifact itself rather than the source record it references.

\begin{figure}[H]
  \centering
  \includegraphics[width=464.32bp]{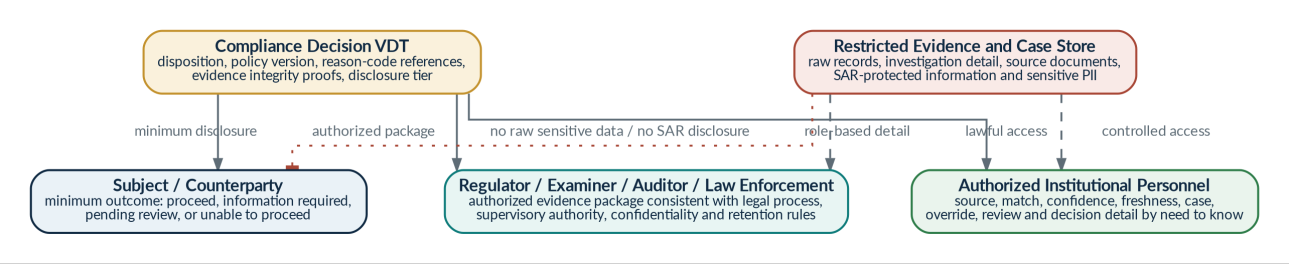}
  \caption{Controlled disclosure of compliance outcomes.}
  \label{fig:disclosure}
  \fignote{The signed decision remains verifiable while raw evidence, investigation detail, sensitive PII, and SAR-protected information remain behind authorized controls.}
\end{figure}

\setlength{\LTleft}{-50.0bp}
\tablepad{12.8bp}{7.2bp}
\begin{longtable}{|B{238.9bp}{8}{10.85}|M{338.5bp}{8}{10.85}|}
\hline
\headrow\thc{Audience} & \thc{Appropriate disclosure}\\\hline
\endfirsthead
\hline
\headrow\thc{Audience} & \thc{Appropriate disclosure}\\\hline
\endhead
Subject or counterparty & Minimum disposition: proceed, additional information required, pending review, or unable to proceed. Do not disclose SAR information, confidential investigative logic, or raw restricted records.\\\hline
\altrow Authorized institutional personnel & Source, match, confidence, freshness, case, analyst, review, override, and decision detail according to role, need to know, and applicable restrictions.\\\hline
Regulator, examiner, auditor, or law enforcement & Authorized evidence package consistent with lawful process, supervisory authority, confidentiality, retention, and privilege requirements.\\\hline
\altrow Payment or technology partner & Only the decision, transaction binding, control completion, status, and proof needed to enforce the relevant session or transaction boundary.\\\hline
\end{longtable}

Raw biometric templates, government-identification images, criminal-history records, SARs and SAR indicators, full sanctions-investigation notes, source-of-funds records, beneficial-ownership documents, and similar sensitive material are not placed in broadly portable \href{https://chainit.com/product/data-tokenization/}{VDTs}. The Compliance Decision VDT contains signed results, evidence references, integrity proofs, reason-code references, and disclosure classification. Underlying evidence remains in restricted stores with access control, retention, correction, redress, and legal-process handling. FinCEN’s SAR confidentiality requirements are an explicit design constraint. \cite{ref9}

\section{Business Value and Measurable Outcomes}\label{sec:value}

\begin{itemize}
  \item Closes the evidence-age gap by evaluating source-specific freshness and context at the moment of reliance rather than treating onboarding as permanent truth.
  \item Produces a session- or transaction-specific, signed, versioned compliance decision instead of a vague assertion that “compliance was checked.”
  \item Keeps authority, compliance, risk tier, session eligibility, transaction policy, capacity, and payment credentials separate so one favorable result cannot silently substitute for another.
  \item Creates a common control boundary across individual-, group-, organization-, and agent-initiated actions, including organization-controlled scheduled-workflow initiations.
  \item Preserves the sovereignty boundary: agents may propose and coordinate, while the deterministic control plane retains final execution authority.
  \item Reduces reconstruction work by linking identity, ARP, delegated authority, evidence, policy, review, approval, reservation, EAC lifecycle, rail outcome, and \href{https://chainit.com/}{Valitorum} artifacts.
  \item Supports selective disclosure by giving each audience only the outcome and evidence detail it is authorized to receive.
  \item Standardizes evidence collection, freshness evaluation, predicate execution, decision recording, and audit production as transaction and agent volume increase.
\end{itemize}

\vspace{-8.4bp}%
\setlength{\LTleft}{-46.6bp}
\tablepad{12.8bp}{7.2bp}
\begin{longtable}{|B{239.3bp}{8}{10.85}|M{331.2bp}{8}{10.85}|}
\hline
\headrow\thc{Outcome} & \thc{Measure}\\\hline
\endfirsthead
\hline
\headrow\thc{Outcome} & \thc{Measure}\\\hline
\endhead
Decision completeness & Percentage of decisions containing required subject, CPP, evidence, freshness, predicate, disposition, reason, scope, expiration, and disclosure fields.\\\hline
\altrow Authority-flow consistency & Percentage of compliance decisions correctly referenced by the corresponding Session Eligibility or Transaction Policy Decision \href{https://chainit.com/product/data-tokenization/}{VDT}.\\\hline
Freshness compliance & Percentage of reliance events decided within applicable evidence-age or status-epoch bounds; count and cause of Stale/Unknown states.\\\hline
\altrow Decision latency & P50, P95, and P99 latency for cached evidence, refreshed evidence, and human-review paths.\\\hline
False positives and false negatives & Valid actions incorrectly held or rejected; actions that should have been escalated or prohibited but were allowed.\\\hline
\altrow Status propagation & Time from authoritative or institutional state change to enforcement by online and cached verifiers.\\\hline
Manual workload & Reviews, overrides, support interventions, and staff time by transaction cohort compared with baseline.\\\hline
\altrow Commit control integrity & Percentage of committed transactions with final Allow, fresh status, capacity reservation, required approvals, matching digest, and valid unused EAC.\\\hline
Audit retrieval & Time to produce a complete independently verifiable identity, authority, evidence, policy, decision, execution, and outcome package.\\\hline
\altrow Privacy and disclosure & Percentage of responses meeting minimum-disclosure policy; unauthorized, excessive, or prohibited disclosure events.\\\hline
Exception continuity & Percentage of later returns, reversals, recalls, chargebacks, and disputes recorded without alteration of the original \href{https://chainit.com/}{Valitorum}.\\\hline
\end{longtable}

\goldpanel{Claims discipline}{Provable Compliance is designed to reduce stale, unsupported, out-of-scope, insufficiently evidenced, and incorrectly disclosed decisions. It is not represented as fixed-cost, universally real-time, fraud-proof, or automatically compliant with every law. Performance, accuracy, propagation, and cost claims should be based on measured deployment data.}

\section{Implementation Priorities and Controlled Deployment}\label{sec:implementation}

The shared \href{https://chainit.com/insights/chainit-launches-provable-authority-as-u-s-generative-ai-enabled-fraud-losses-are-projected-to-reach-40-billion-by-2027/}{Authority Protocol} definitions and flows are controlled by CIT-PA-WP-7.0. The compliance implementation should therefore be developed as a companion layer rather than as a parallel authority system. The immediate work should preserve that division and validate the evidence, policy, decision, disclosure, and lifecycle interfaces described here.

\setlength{\LTleft}{-53.8bp}
\tablepad{12.8bp}{7.2bp}
\begin{longtable}{|B{239.2bp}{8}{10.85}|M{345.5bp}{8}{10.85}|}
\hline
\headrow\thc{Priority} & \thc{Required outcome}\\\hline
\endfirsthead
\hline
\headrow\thc{Priority} & \thc{Required outcome}\\\hline
\endhead
1. Canonical schemas & Freeze Evidence Assertion \href{https://chainit.com/product/data-tokenization/}{VDT}, CPP, BRE predicate, Compliance Decision VDT, review-case, disclosure-tier, and status/freshness schemas.\\\hline
\altrow 2. Authority interface & Define the exact Compliance Decision Permits interface to Session Eligibility Decision \href{https://chainit.com/product/data-tokenization/}{VDT} and Transaction Policy Decision VDT; prohibit direct EAC issuance from a compliance result.\\\hline
3. Subject model & Implement individual, group/quorum, and organizational representative inputs, including organization-controlled scheduled-workflow initiations, together with Agent Subject Profile inputs, without creating conflicting authority semantics.\\\hline
\altrow 4. Evidence-provider inventory & Document source class, fields, permitted purpose, coverage, confidence, latency, cost, refresh model, dispute path, outage behavior, and retention for each provider.\\\hline
5. Policy and workflow governance & Version CPPs; separate predicate evaluation, risk-tier assignment, disposition mapping, manual review, override authority, and policy approval.\\\hline
\altrow 6. Freshness and status & Define source-specific maximum ages, event-driven invalidation, status epochs, live-verification triggers, degraded-mode behavior, and propagation measurements.\\\hline
7. Privacy and disclosure & Implement audience-specific responses, restricted evidence stores, SAR confidentiality, selective disclosure, correction/redress, and access audit.\\\hline
\altrow 8. Lifecycle integration & Link compliance decisions to the exact session or transaction artifacts, Execution Authorization Lifecycle \href{https://chainit.com/product/data-tokenization/}{VDTs}, \href{https://chainit.com/}{Valitorum}, and Supplemental Valitorum.\\\hline
9. Conformance and adversarial testing & Test stale evidence, source outages, false matches, changed digests, concurrent requests, incomplete controls, improper EAC issuance, unauthorized disclosure, and later exceptions.\\\hline
\altrow 10. Patent and publication review & Reconcile the final implementation against filed support and complete any required priority or new-matter work before external technical disclosure.\\\hline
\end{longtable}

\subsection{Open implementation questions}\label{sec:questions}

\begin{itemize}
  \item Which evidence providers and authoritative-source interfaces are contracted, available by jurisdiction, and permitted for each purpose?
  \item Which policy-profile templates will ChainIT provide, and which logic remains solely institution-configured?
  \item How will risk-tier assignment, manual override, and policy approval be governed and independently tested?
  \item What status-epoch, live-query, and evidence-refresh service levels can be supported for each source and partner?
  \item How will false positives, disputes, corrections, consumer notices, and adverse actions be routed and evidenced?
  \item What data may be disclosed to subjects, counterparties, payment partners, auditors, regulators, and law enforcement?
  \item Which session and transaction artifacts are currently implemented, in pilot, or proposed, and how will that status be represented externally?
  \item How will rail-specific execution and later exceptions be linked to the compliance record without altering the original \href{https://chainit.com/}{Valitorum}?
\end{itemize}

\section{Related Work}\label{sec:related}

\runin{Verifiable credentials and status.} The W3C Verifiable Credentials Data Model \cite{vcdm2} standardizes cryptographically verifiable claims, and the Bitstring Status List \cite{bsl} and OAuth~2.0 Token Introspection \cite{rfc7662} provide cached and live status, respectively. Evidence Assertion \href{https://chainit.com/product/data-tokenization/}{VDTs} apply the same separation between a signed statement and its current status, but treat each sourced fact as a separate, purpose-limited assertion with a source-specific freshness bound rather than as a credential of standing eligibility (Sections~\ref{sec:evidence} and~\ref{sec:runtime}).

\runin{Policy-based decisioning.} Attribute-based access control \cite{abac} and policy languages such as XACML \cite{xacml} evaluate attributes of subjects, resources, and environment against versioned policies. The Compliance Policy Profile and BRE predicates follow this pattern; the result, however, is a signed, time-bounded Compliance Decision \href{https://chainit.com/product/data-tokenization/}{VDT} that records which evidence and policy version produced it, and it is consumed as one predicate by the \href{https://chainit.com/insights/chainit-launches-provable-authority-as-u-s-generative-ai-enabled-fraud-losses-are-projected-to-reach-40-billion-by-2027/}{Authority Protocol} rather than serving as an access decision in itself (Sections~\ref{sec:cpp}–\ref{sec:input}).

\runin{Regulatory technology.} Research on regulatory technology (RegTech) describes the use of technology to automate monitoring, reporting, and compliance in financial services \cite{arner2017}. \href{https://chainit.com/white-papers/}{Provable Compliance} does not replace an institution’s AML/CFT or sanctions program (Section~\ref{sec:problem}); it addresses how the evidence, policy selection, and disposition behind each decision can be made attributable and independently verifiable at the moment of reliance.

\runin{Delegated authority and tamper-evident records.} Recent work on authenticated delegation to AI agents \cite{south2025} and the companion \href{https://chainit.com/insights/chainit-launches-provable-authority-as-u-s-generative-ai-enabled-fraud-losses-are-projected-to-reach-40-billion-by-2027/}{Authority Protocol} \cite{ref23} address who may act; this paper supplies the compliance-side predicate consumed by that protocol. Append-only lifecycle \href{https://chainit.com/product/data-tokenization/}{VDTs} and the sealed \href{https://chainit.com/}{Valitorum} relate to tamper-evident logging \cite{crosby2009}.

\section{Conclusion}\label{sec:conclusion}

The most important compliance question is not whether a check occurred once. It is whether the evidence and rule set required for the present reliance event are current enough, applicable, and sufficiently certain for the organization’s configured policy. That question cannot be answered by a universal badge because sanctions, PEP status, AML risk, background information, identity assurance, licensing, KYC/KYB, and payment-transparency obligations have different sources, volatility, legal effects, and disclosure constraints.

ChainIT \href{https://chainit.com/white-papers/}{Provable Compliance} turns those differences into a verifiable compliance layer. \href{https://chainit.com/product/identity-verification/}{ChainIT ID} and \href{https://chainit.com/product/business-verification/}{ChainIT Org ID} establish the subjects. Evidence Assertion \href{https://chainit.com/product/data-tokenization/}{VDTs} preserve sourced and institution-generated facts with provenance and freshness. A versioned CPP selects the rules. The BRE returns deterministic, reason-coded predicates. The organization’s Workflow Engine assigns risk tier where applicable and composes the predicates into a signed, time-bounded Compliance Decision VDT. That decision gates present exercise without rewriting the ARP or creating authority.

The final \href{https://chainit.com/insights/chainit-launches-provable-authority-as-u-s-generative-ai-enabled-fraud-losses-are-projected-to-reach-40-billion-by-2027/}{Authority Protocol} remains controlling. The Compliance Decision \href{https://chainit.com/product/data-tokenization/}{VDT} supplies the Compliance Decision Permits predicate to the Session Eligibility Decision VDT or Transaction Policy Decision VDT. A SAC enables proposal activity but is never payment authority. A transaction can commit only after final Allow, fresh status, exact-digest matching, required approvals, atomic capacity reservation, and a valid unused EAC. Execution Authorization Lifecycle VDTs preserve credential treatment, the \href{https://chainit.com/}{Valitorum} seals the terminal evidence graph, and Supplemental Valitorums preserve later exceptions without modifying history. \cite{ref23}

The resulting control is neither “trust the onboarding file” nor “query every database for every action.” It is a disciplined point-of-reliance decision: rely on signed evidence while it remains fresh enough; refresh when source, risk, or policy requires; preserve uncertainty as Review, Hold, or Stale/Unknown; apply prohibitions only to the scope they actually govern; disclose only what the audience may receive; and supply the final \href{https://chainit.com/insights/chainit-launches-provable-authority-as-u-s-generative-ai-enabled-fraud-losses-are-projected-to-reach-40-billion-by-2027/}{Authority Protocol} with a verifiable compliance predicate before consequence. That is the zero-trust compliance foundation for scalable human, organizational, and agentic commerce.

\unnumbered{References and Source Basis}

{\fontsize{9bp}{12.2bp}\selectfont\color{tgray}Regulatory references are stated as current through August 31, 2026 based on the official sources listed in this edition. Each source should be revalidated immediately before publication and before deployment-specific reliance. ChainIT patent and application materials support the identity, authority, \href{https://chainit.com/product/data-tokenization/}{VDT}, BRE, \href{https://www.pactvera.com/}{Pactvera}, conditional-execution, and \href{https://chainit.com/}{Valitorum} foundations described here; CIT-PA-WP-7.0 controls the shared \href{https://chainit.com/insights/chainit-launches-provable-authority-as-u-s-generative-ai-enabled-fraud-losses-are-projected-to-reach-40-billion-by-2027/}{Authority Protocol} definitions and flows. \cite{ref17}–[\citen{ref23}]\par}

\begingroup
\fontsize{9.5bp}{12.8bp}\selectfont
\makeatletter
\renewcommand{\@biblabel}[1]{\textbf{\color{cnavy}[#1]}}
\renewenvironment{thebibliography}[1]
  {\list{\@biblabel{\@arabic\c@enumiv}}{\leftmargin0bp\labelwidth0bp\labelsep0bp\itemindent0bp\listparindent0bp
     \itemsep6bp\parsep0bp\topsep0bp\partopsep0bp\usecounter{enumiv}\renewcommand{\makelabel}[1]{##1\ }}\RaggedRight}
  {\endlist}
\makeatother

\endgroup

\end{document}